\documentclass[twocolumn]{emulateapj}

\def\alwaysmath#1{\ifmmode{#1}\else{$#1$}\fi}
 \slugcomment{Accepted to {\it
The Astrophysical Journal}.}
\shortauthors{MEN\'{E}NDEZ-DELMESTRE ET AL.}
\shorttitle{2MASS BAR FRACTION \& BAR PROPERTIES}

\begin{document}

\title{A Near-Infrared Study of 2MASS Bars in Local Galaxies: An Anchor for High Redshift Studies} \author{\sc Kar\'{i}n
Men\'{e}ndez-Delmestre\altaffilmark{1}, Kartik
Sheth\altaffilmark{1,2}, Eva Schinnerer\altaffilmark{3}, Thomas
H. Jarrett\altaffilmark{4}, Nick Z. Scoville\altaffilmark{1}}

\altaffiltext{1}{California Institute of Technology, Department of
Astronomy, Mail Stop 105-24, Pasadena, CA 91125}

\altaffiltext{2}{Spitzer Science Center, Mail Stop 220-6, California
Institute of Technology, Pasadena, CA 91125}

\altaffiltext{3}{Max-Planck-Institute for Astronomy, 69117 Heidelberg,
Germany}

\altaffiltext{4}{Infrared Processing and Analysis Center, 100-22,
California Institute of Technology, 770 South Wilson Avenue, Pasadena,
CA,91125}

\email{km@astro.caltech.edu}

\begin{abstract}

We have measured the fraction of bars in nearby spiral galaxies using
near-infrared $J$, $H$, and $K_s$ images of 151 spiral galaxies from
2MASS.  This local sample provides an anchor for the study of the
evolution of the bar fraction and bar properties with redshift. We
identify bars by analyzing the full two-dimensional light distribution
and requiring a combined ellipticity and position angle signature. The
combined \textit{bar signature} is found in 59\% of the galaxies.  The
bar fraction increases to 67\% when we include ``candidate'' bars,
where only the ellipticity signature is present. We also measure the
change in the bar fraction as a function of bar size; the bar fraction
drops to 36\% for bars with a semi-major axis larger than 4\,kpc. We
find that infrared bars typically extend to one-third of the galactic
disk, with a deprojected relative size of $<a_{bar}/R_{25}> \sim 0.3
\pm 0.2$. Early-type spirals host significantly larger bars,
consistent with earlier studies. The $<a_{bar}/R_{25}>$ is two times
larger in early-types than in late-types. The typical bar axial ratio
(b/a) is $\sim$0.5, with a weak trend of higher axial ratios for
larger bars.
\end{abstract}


\section{INTRODUCTION}

Bars play a central role in the evolution of galaxies. The
non-axisymmetry of the bar induces large-scale streaming motions in
the stars and the gas \citep{a92a, piner95, teuben95}.  Unlike stars,
the gas is collisional and dissipative, losing angular momentum and
flowing inwards down the bar dust lanes \citep{combes85, a92b,
  regan97b, regan99, sheth00, sheth02, sheth05}. This inflow leads to
dramatic changes in the host galaxy such as accumulation of molecular
gas in the central kiloparsec \citep{sakamoto99, sheth05}, smoothing
of the chemical abundance gradient \citep{martin94}, inducement of
circumnuclear star formation (see \citealt{ho97a, ho97b} and
references therein) and (possibly) the formation of bulges and
pseudo-bulges \citep{norman96, kormendy04, sheth05}, fueling of active
galactic nuclei \citep{shlosman89}, and perhaps the destruction of the
bar itself \citep{norman96, das03}.  Understanding the bar fraction
and bar properties is therefore critical to understanding the
evolution of spiral galaxies.

As early as 1963, using photographic plates that are sensitive to blue
light, de Vaucouleurs \citep{devau63} found that
35\% of all nearby spiral galaxies (S0/a-Sd) are strongly barred (SB).
An additional 29\% were classified as ``intermediately''-barred (SAB)
spirals. With the advent of near-infared (NIR) detectors, previously
undiscovered bars were seen at infrared wavelengths (e.g.,
\citealt{hackwell83, scoville88, thronson89, mulchaey97, seigar98,
  jarrett03}); these were typically small nuclear bars at the centers
of nearby spirals, where dust obscuration can be high and variable.
In hindsight these discoveries are not surprising since bars are
stellar structures that are best seen at longer wavelengths, where the
veiling effects of dust extinction and star formation are minimized
compared to optical wavelengths.  The NIR light is dominated by old
stars which constitute the bulk of the stellar mass; the NIR, therefore,
is a more reliable tracer of a galaxy's gravitational potential than
optical light. Although large format NIR cameras have become
more common over the last decade, there are only a few large and
homogeneous NIR surveys of nearby spirals, and of these, only the Ohio
State University Bright Spiral Galaxy Survey (OSUBSGS;
\citealt{eskridge02}), with 205 spirals with T\,$\geq\,0$, B
$\leq\,12$ and D$\leq\,6 \arcmin$ has been used to examine the
fraction of bars in the H-band (\citealt{eskridge00, whyte02,
  laurikainen04}, hereafter E2000, W2002 and LSB04,
respectively). Several other studies have examined the bar
fraction in the optical and NIR between active and non-active
galaxies, to determine the role of bars in feeding AGN
\citep{mulchaey97, knapen00, laine02, laurikainen04}.

The advent of large, deep extragalactic surveys such as COSMOS, GOODS, and 
GEMS \citep{scoville06, dickinson01, rix04} has triggered studies
that explore the evolution of the bar fraction with redshift
\citep{sheth03, sheth04, elmegreen04, jogee04}. A prerequisite for 
evaluating the results of these high redshift studies is a well-determined bar
fraction in the local universe.  

Using H-band imaging of 186 and 113 nearby galaxies respectively,
E2000 and W2002 classified $\sim 3/4$ of the spirals as barred. E2000
relied on visual inspection to identify bars, like earlier optical
studies (e.g., Third Reference Catalog of Bright Galaxies,
\citealt{devau91}, hereafter, RC3), while W2002 relied on an automated
method of measuring a bar based on the difference in the axial ratio
and position angles of a best-fit ellipse to one interior and one
exterior isophote; this method is a refinement of the bar identification method used by \citet{abraham96}.

Although the eye is an excellent classification tool, visual
inspection of images is unreliable for poor quality and low
signal-to-noise data.  Especially in the context of high redshift
galaxies, visual classification quickly becomes difficult in view of
the significant decline in the spatial resolution (e.g. Figure\,3 of
\citealt{sheth03}) or decreased signal to noise (due to the cosmological surface
brightness dimming) with increasing redshift. Classification by eye is
also tedious for large data sets and is subjective when comparing
datasets of varying quality.  In such cases an automated method for
bar identification is more useful as it has the advantage of
reproducibility and can be applied to large datasets.  Moreover, an
automated algorithm that uses the full two-dimensional light
distribution is likely more robust than one that relies on a small
number of isophotes.  

An accurate measurement of the bar fraction also depends on a variety
of selection biases such as surface brightness limits, signal to noise
ratio, inclination and spatial resolution \citep{sheth03, sheth04}.
The measurement of the bar fraction is particularly sensitive to the
spatial resolution of the data. \citet{sheth03} show how the bar
fraction measured in the Hubble Deep Field North
using the coarse NICMOS data is comparable to the local RC3 fraction
when the bar size is taken into account.  It is, therefore, critical
to have accurate measurement of the bar size distribution.

A number of studies have looked at the bar size distribution of
optically-selected (i.e. RC3 SB/SAB) local bars \citep{kormendy79,
  elmegreen85, martin95, erwin05}.  These optical studies have found
that bars in early-type spirals tend to be longer than those found in
late-types.  \citet{laine02} and LSB04 have compared properties of
H-band selected bars in active and non-active galaxies.

In this paper we present a detailed NIR morphological
analysis of the local spiral population imaged by the Two Micron All
Sky Survey \citep{skrutskie06} Large Galaxy Atlas (hereafter, 2MASS
LGA, \citealt{jarrett03}). We identify bars and characterize their
sizes and ellipticities with a widely-used technique of fitting
ellipses (e.g., \citealt{wozniak91, regan97a, zheng05} in the optical;
\citealt{knapen00}, \citealt{laine02}, \citealt{laurikainen02} and \citealt{sheth00,
  sheth02, sheth03} in the NIR) to the full two-dimensional light
distribution of the spirals in our 2MASS sample.

2MASS is one of the largest and most homogeneous NIR surveys.
It allows us to measure the local bar fraction
and characterize  the bar properties and their relationship to their disks.  This work
provides a local anchor for studies of the evolution of the bar
fraction with redshift.  

\begin{figure}
\plottwo{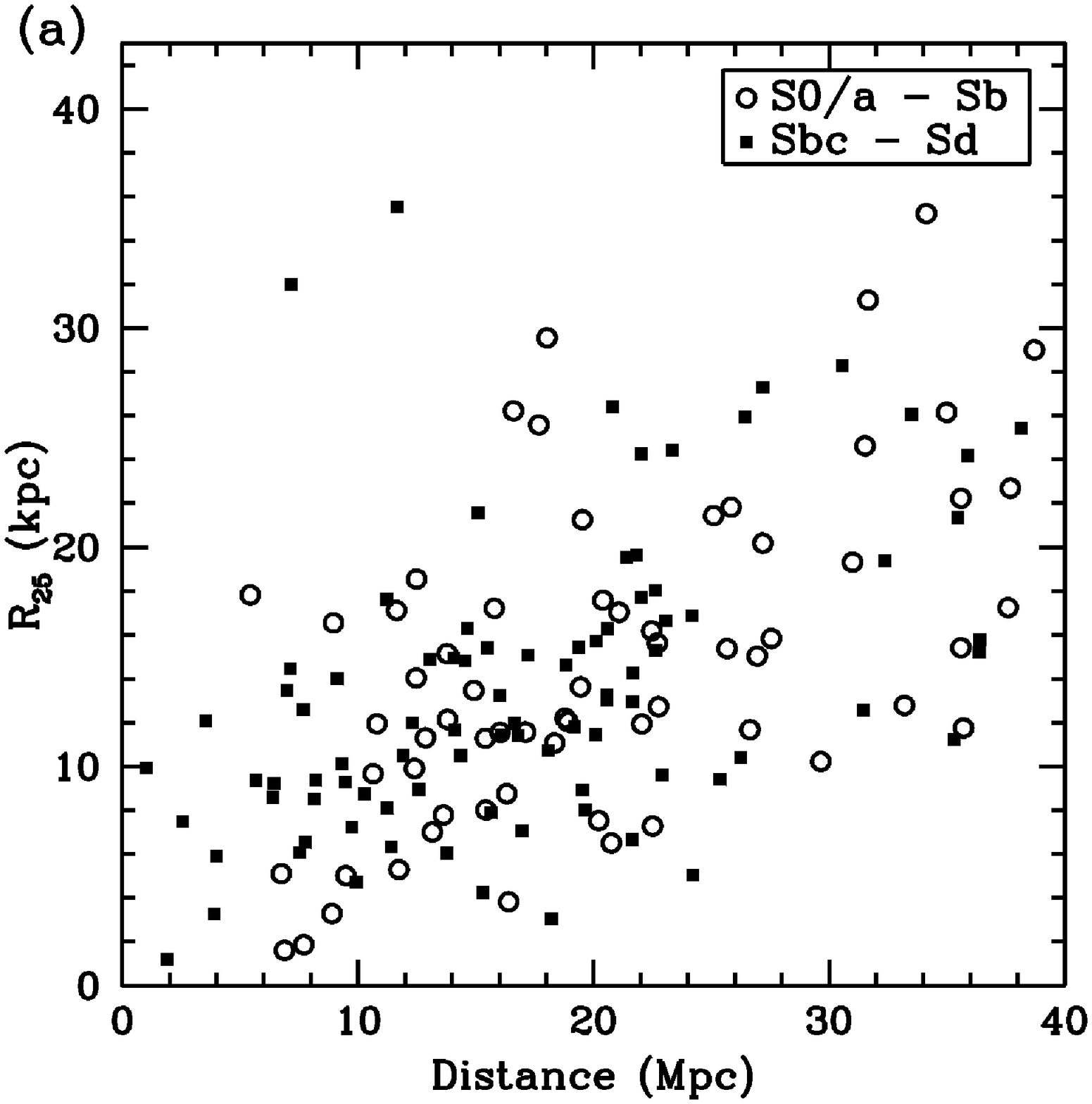}{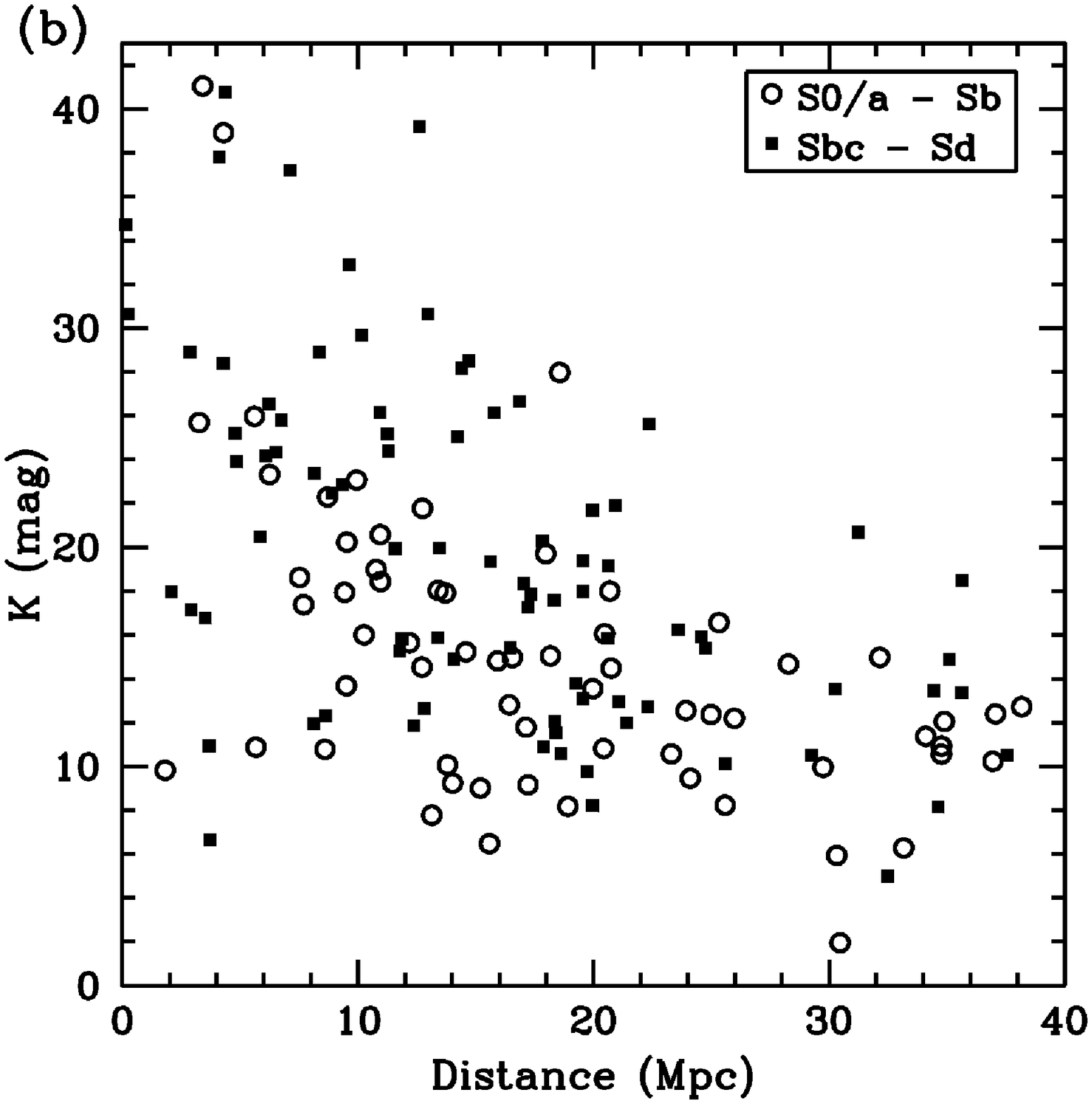}
\caption{(a) Galaxy optical radius as a function of distance for our 2MASS spiral sample.  (b) Same plot for absolute $K$-band magnitude. Early-type (S0/a-Sb) and late-type (Sbc-Sd) spirals are shown as open circles and solid squares, respectively.}
\end{figure}

\section{OBSERVATIONS and ANALYSIS}

\subsection{Defining a Local Sample}

2MASS imaged the entire sky in the NIR bands, $J$ ($1.2\, \mu$m), $H$
($1.6\, \mu$m), and $K_s$ ($2.2\, \mu$m) down to a limiting sensitivity of
21.6, 20.6 and 20.0 mag/arcsec$^2$ ($1 \sigma$), respectively,
with a typical angular resolution of $\sim 2-3 \arcsec$. From the
image data set, which contains galaxies with $K_s \textless 14$ mag,
\citet{jarrett03} assembled the LGA.  The LGA consists of individual
and co-added (to improve signal-to-noise) $J$, $H$ and $K_s$ band
images of over five hundred large galaxies imaged in the 2MASS, with
sizes ranging from $2 \arcmin$ to $2^\circ$.  

In the absence of an established NIR morphological classification system, we used the optical RC3 morphological classification for our sample selection.  We selected all 339 LGA galaxies with good signal-to-noise that were
identified as spirals (S0/a-Sd) in the RC3.  From these, we
excluded 165 highly inclined galaxies ($i > 65^\circ$, where inclination
is derived from the disk axial ratio).   We excluded galaxies classified as irregulars, mergers or otherwise strongly interacting.  We also limited our sample to
the 151 closest galaxies ($D  \textless  40$ Mpc) so that we may easily resolve
potential bars.  Distances to these spirals were taken from the NASA Extragalactic Database (NED). The median distance of our sample of galaxies is 18 Mpc. There is an inherent bias towards large, massive and bright galaxies in 2MASS because it is relatively shallow (see Figure\,1).  The list of galaxies
selected and their global properties are listed in Table
\ref{sampletab}.

\subsection{The Bar Signature}\label{signature}

We detect the presence of a bar within a galaxy disk from the
ellipticity and position angle (hereafter, PA) profiles traced by the
galaxy isophotes.  We applied the IRAF task $ellipse$
\citep{jedrzejewski87} on the combined $J+H+K_s$ LGA images for all
the galaxies in our sample, with a step size of $3 \arcsec$ (the 2MASS
angular resolution).  The transition from the (round) bulge-dominated
center to the disk is typically characterized by an increase in
ellipticity, $\epsilon = 1 - (b/a)$.  For a disk or a bar, the PA
remains constant unless there are spiral arms, which force the PA to
change continuously.  In absence of these arms, the ellipticity
increases monotonically at a constant PA within the bar region. At the
end of the bar, the ellipticity drops abruptly and the PA changes
sharply as the isophotes transition from the bar into the disk.  An
example of what we term an ideal \textit{bar signature} is shown in
Figure\,2 for NGC\,1300. In this case, within the bar region, the bar
isophotes show a continuous increase in ellipticity, maintaining a
constant PA.  At the end of the bar, the ellipticity of the image
isophotes decreases and the PA changes sharply, denoting the
difference between the bar and the disk PA.  Cases in which the bar
and the disk have the same PA are discussed below.

\begin{figure}
\plotone{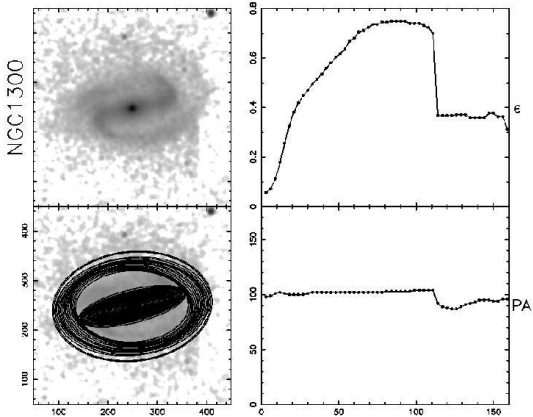}
\caption{top left: 2MASS $J+H+K_{s}$ combined image of NGC\,1300. lower left: NIR image of NGC\,1300 with ellipse-fit ellipses overplotted. top right: ellipticity profile for NGC\,1300. lower right:  PA profile for NGC\,1300.  This is the bar signature in NGC\,1300: $\epsilon$ increases monotonically and drops abruptly with $\Delta\epsilon>0.1$, as ellipses conform to the disk isophotes. The position angle stays constant within the bar region and changes, with $\Delta$PA $>10^\circ$, when the end of the bar is reached.}
\end{figure}

The change in ellipticity and the change in PA are moderated by the
spiral arms that may retard the corresponding PA drop to occur
further out in the disk, outside the bar region. For this reason we
choose the semi-major axis (sma) of the isophote maximum ellipticity ($\epsilon_{max}$)
to be the size of the bar, $a_{bar}$. Due to spiral arms, the
ellipticity peak is often broadened into a ``flat-top'' (see Figure
2). To account for the uncertainty in defining which sma corresponds to the maximum ellipticity, we characterize the error in the bar size
measurement by the sma range encompassing the tip of the ellipticity
peak in the bar signature, over which $\epsilon \gtrsim
(\epsilon_{max} - \delta \epsilon)$, where we chose $\delta
\epsilon = 0.01$.

In order to automate our bar identification method, we required a bar
to have a projected ellipticity, $\epsilon_{max}$, greater than $0.2$,
and the end of the bar to be marked by a change in ellipticity,
$\Delta\epsilon > 0.1$, with an accompanying change in the position
angle, $\Delta$PA $> 10^\circ$.  We do not identify a bar signature in
the first three points of the ellipticity profile, which
corresponds to 3 times the PSF FWHM ($9 \arcsec$).  More than 95\% of all bars in
our sample are longer than 5 times the PSF FWHM.  Therefore, a more
realistic threshold to our bar detection method for identifying bars
is $15 \arcsec$, which corresponds to 1.3\,kpc at the median distance
of 18 Mpc in our sample. This size scale makes this study an ideal
comparison sample for high redshift galaxies where similar spatial
resolution is typically achieved.  We discuss this issue in more
detail in \S \ref{barsizes}.  Our reason for choosing $\epsilon > 0.2$
is to distinguish between bars and oval structures (e.g., flattened
bulges or inner disks).  Our choices for the $\Delta \epsilon$,
$\Delta$PA are based on previous studies which have used these
criteria to identify bars \citep{knapen00, laine02,sheth03}.

Some of the galaxies in our sample have ellipticity profiles that
conform to the bar signature described above, but have PA profiles
that do not\footnote{In our sample, $15\%$ (22/151) of all spirals
  fell into this category.}.  Visual inspection of these images shows
that some appear to contain a bar, while others do not. These cases
require special consideration as discussed here and illustrated in
Figure\,3a--e.  The PA profiles of these galaxies deviate from the PA
bar signature in one of two ways: (1) the change in PA that
accompanies the drop in ellipticity is less than $10^\circ$, or (2)
the PA varies continuously within the region where there is a
monotonic increase in ellipticity.

\begin{figure*}
\plotone{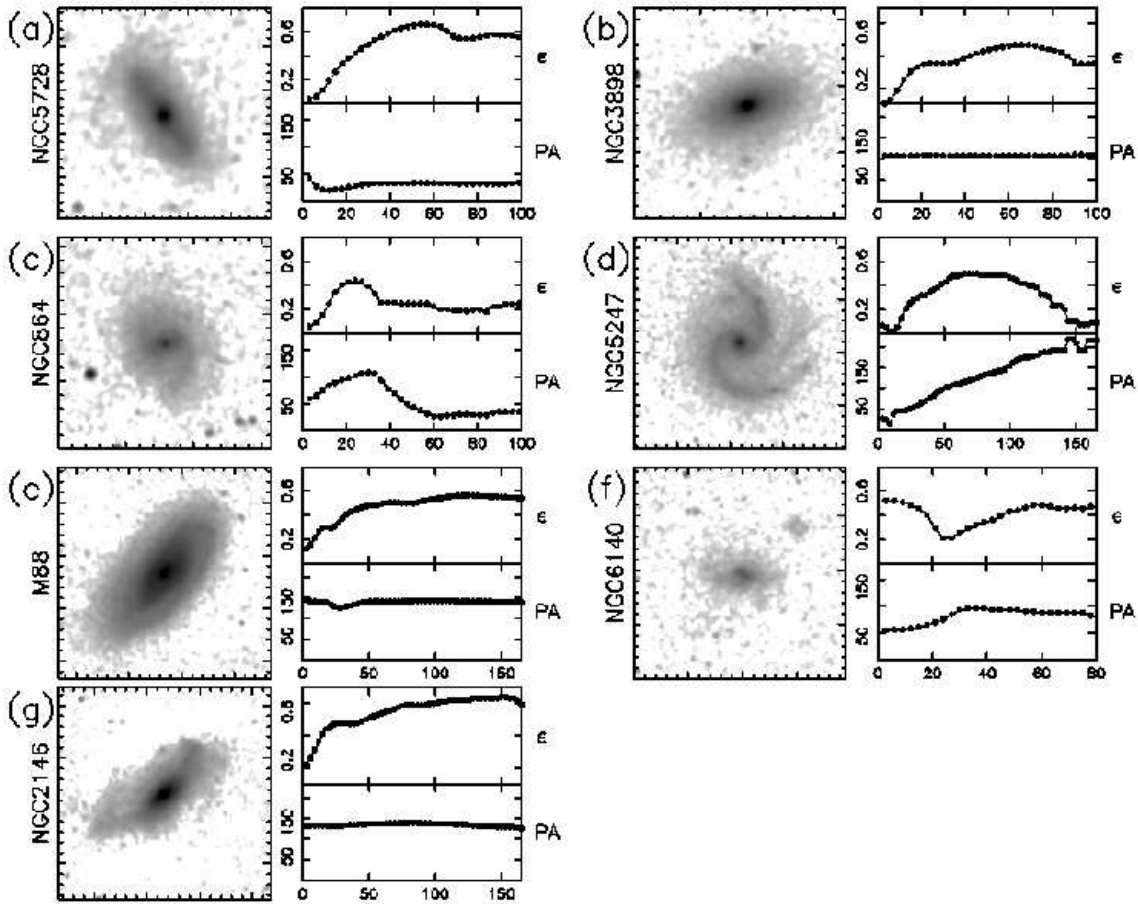}
\caption{Ellipse-fit results following the same format as Figure\,2, excluding image with ellipses overplotted. (a) NGC\,5728 is classified as a candidate, but contains a bar with PA equal to that of the disk. (b) NGC\,3898 displays a PA profile similar to that of candidate NGC\,5728, but here it corresponds to a bright inner disk: it is not barred. (c) NGC\,864 is a candidate that contains a bar, even though its PA profile displays a continuous change in PA within the bar region.  This is due to the presence of spiral arms, which result in the twisting of the bar isophotes within the bar region. (d) NGC\,5247 is an example of a candidate in which the presence of spiral arms may mimick a bar ellipticity signature: it is not barred. (e) M88 is an unbarred galaxy that shows no sign of bar presence in its ellipticity profile. (f) NGC\,6140 is a spiral classified as strongly barred (SB) in the RC3, whose bar is smaller than our detection limit of 15 $\arcsec$.  Ellipse-fit of its 2MASS NIR image does not trace the ellipticity increase associated to the bar. (g) NGC\,2146 is another SB spiral (RC3) for which our method fails to recognize its bar. Following the ellipticity monotonic increase characteristic of a bar signature, instead of dropping sharply, the ellipticity of the galaxy increases further due to the presence of an open-angle spiral arm.}
\end{figure*}

There are two possibilities for the first case: (a) a bar is present,
but it has a PA similar to that of the disk (e.g. Figure\,3a), or (b)
there is no bar and the bar-like ellipticity  signature traces the
presence of an inner disk embedded within the galaxy disk (e.g. Figure
3b). Since bars are expected to be randomly oriented with respect to
the underlying disk and $\Delta$PA may vary within $0^\circ\,\lesssim\,\Delta\,PA\,\lesssim\,90^\circ$; we expect that, based strictly on
geometric considerations, $\sim\,10\%$ of bars will have a PA within
$10^\circ$ of the host galaxy disk PA.

In the second case, the PA does not remain constant within the region
where the ellipticity increases monotonically, but varies
continuously. Two scenarios are responsible for this behavior, both
involving the presence of spiral arms: (a) spiral arms originating
within the bar region twist the bar isophotes, producing a progressive
change in PA like the example shown in Figure\,3c; or (b) no bar is
present but the ellipticity profile is instead produced by the
progressive stretching of disk isophotes by fairly open, bright spiral
arms. This is shown in Figure\,3d.

If we were to consider solely the ellipticity signature, all of the
cases with these different PA signatures would be classified as barred
galaxies, even though they probably do not all host bars.  Instead, we
adopt a separate category for these galaxies, introduced by
\citet{sheth03}: we classify these galaxies as $candidate$ barred
spirals and will hereafter be referred to as $candidates$. We visually
inspected each individual candidate to determine whether a real bar
was present or not.  Candidates for which the ellipticity signature is
produced by spiral arms (Figure\,3d) or an inner disk (Figure\,3b) are
classified as $unbarred$.  The remaining candidates are classified as
$barred$.

Galaxies that do not show a monotonic rise and fall in ellipticity are classified as $unbarred$ galaxies.  An example
of an unbarred galaxy is shown in Figure\,3e, the nearby spiral, M88.  

\section{RESULTS}

\subsection{Bar Fraction}\label{fractionres}

We successfully applied the ellipse-fit technique to the 151 galaxies
in our sample. We found that 89 (59\%) of all galaxies in our sample
display both an ellipticity and PA bar signature, and are therefore
identified $clearly$ as barred spirals.  This fraction represents a
lower limit to the NIR bar fraction measured in our 2MASS
sample.  Of the candidate barred spirals, where only one of the two
bar signatures is present, 12/22 appear to have bars upon visual
inspection.  Thus, the fraction of galaxies with bars in our 2MASS
sample increases to 101/151 (67\%).  For the remainder of the paper,
we refer to this 67\% sample as \textit{barred spirals}.  Our
classification and notes for each galaxy, as well as their RC3
classifications are presented in Table \ref{resultstab}.

We identify 86\% of all SB galaxies and 80\% of all SAB galaxies from
the RC3 as barred spirals (see Table \ref{resultstab}). In addition,
we classify 11 galaxies within the RC3 SA category as barred
spirals. With our exclusion of highly inclined galaxies ($i >
65^\circ$), our sample has 64 galaxies in common with those of
E2000. Our classification agrees in 85\% of the cases with the visual
classification by E2000. Most of the RC3 and E2000 SB and SAB
galaxies that do not meet our bar identification requirement have
either bars that are too small, or have insufficient change in the
ellipticities due to open spiral arms and/or disks with higher
ellipticites than the bar; two such examples are NGC\,6140 and NGC\,2146
shown in Figures 3f and 3g.

Following our adopted definition for the bar signature, there are a
number of instances in which our method fails to identify bars (see
\S\ref{signature}). Our measurement of the bar fraction is thus a firm
lower limit to the true value, but it is not straightforward to
quantify the error in the local bar fraction.  Poisson statistics
would indicate an error of $\pm$2\% on the total bar fraction of 67\%.

\subsection{Bar Properties}

\begin{figure}
\plotone{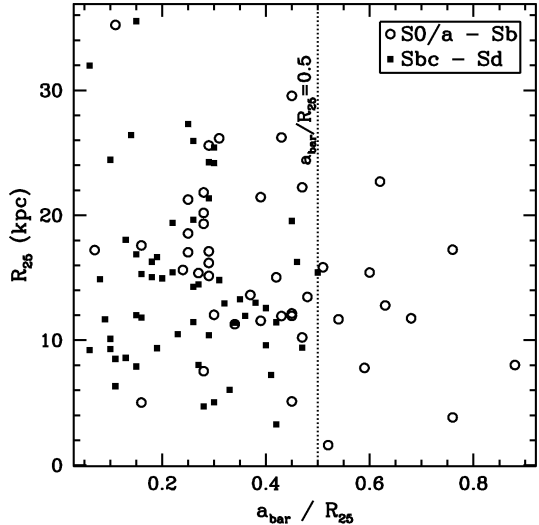}
\caption{$a_{bar} / R_{25}$ as a function of $R_{25}$. Symbols for early-type and late-type spirals follow the same convention as Figure\,1. Note that $\sim90\%$ of all bars extend out to less than half of the underlying disk radii and the bars largest with respect to the blue disk of their host galaxy are mostly found in the smaller early-type galaxies. }
\end{figure}

\subsubsection{Bar Sizes}\label{barsizes}

We define the relative bar size, $a_{bar} / R_{25}$, as the ratio of
the bar deprojected semi-major axis to the RC3 radius of the host
galaxy at a B magnitude of 25 ($R_{25}$).  The bar relative size has the
advantage of being a distance-independent measure.  The error on the
relative bar size is $\sim 10\%$ and is dominated by the uncertainty
on the measurement from the ellipticity profile of the bar
signature. We found that $\sim$90\% of all bars within the sample
extend out to less than 50\% of their host galaxy disk. Bars with the
largest relative sizes are mostly found in small early-type galaxies
(see Figure\,4) with $R_{25} \textless 23$\,kpc. The properties of the
typical 2MASS bar are summarized in Table \ref{TypBartab}.

\begin{figure*}
\plottwo{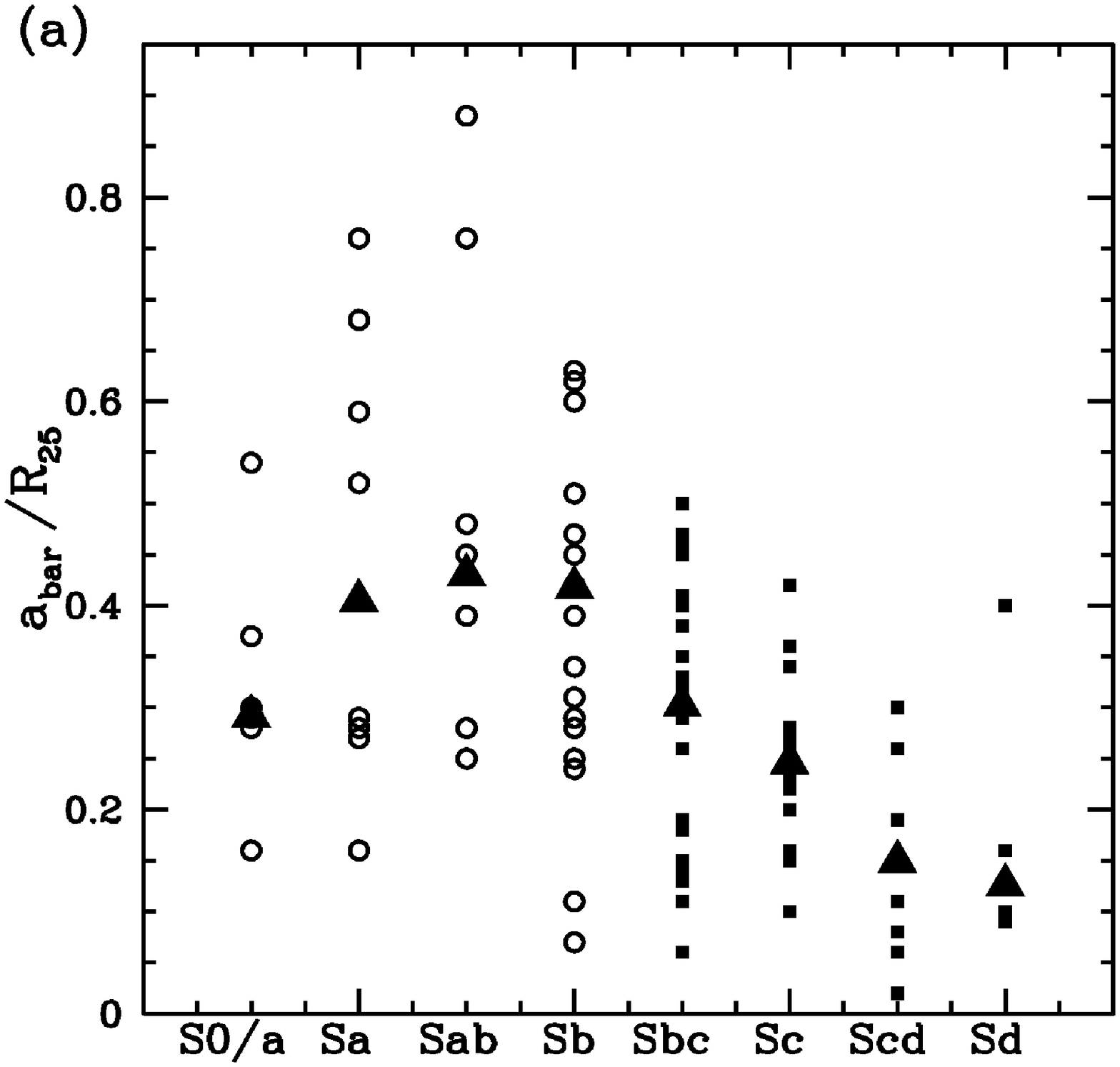}{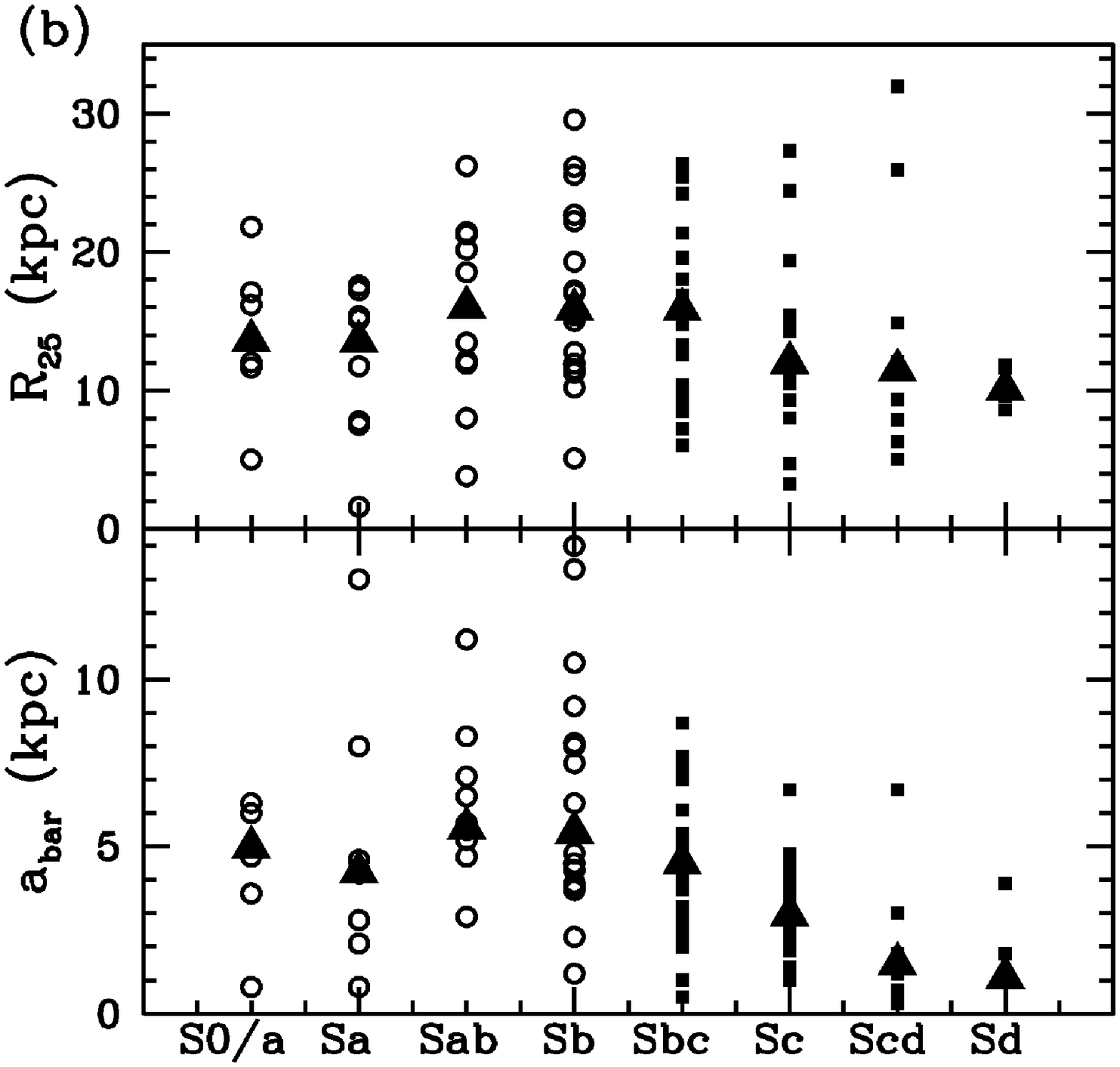}
\caption{(a) $a_{bar} / R_{25}$ as a function of Hubble Type. Symbols for early-type and late-type spirals follow the same convention as Figure\,1. Solid triangles show median values for each Hubble Type. A clear trend shows that early-type bars tend to be generally larger than late-type (Sbc-Sd) bars. S0/a bars have a smaller median size than that expected from the trend of bar sizes along the Hubble sequence.  However, this is likely due to the low number of objects that we have in this category. (b) Same as (a), but for $R_{25}$ (top) and $a_{bar}$ (bottom).  Note that both $a_{bar}$ and $R_{25}$ tend to be smaller for late-type spirals.  This confirms that the trend of smaller relative sizes in late-type spirals, shown in (a), is driven by bars being intrinsically smaller in late-type spirals than in early-types.}
\end{figure*}

We find a clear trend of bar sizes, both in terms of $a_{bar}$ and
$a_{bar}/R_{25}$, along the Hubble sequence, with early type bars
being significantly larger than late types (see Figure\,5).  Bars in
early Sa-Sb spirals have a median $a_{bar}/R_{25} = 0.43 \pm 0.04$,
while bars in later-type Sc-Sd spirals have a median $a_{bar}/R_{25} =
0.2 \pm 0.02$, making the median relative bar size in early-type
galaxies larger than in late types by a factor of two. Note that the
low median value for the S0/a spirals is probably due to the low
number of these transitional lenticular-to-spiral galaxies in our
sample. Furthermore, the sizes of early-type bars cover a large range
in relative sizes, $0.1 \textless a_{bar} / R_{25} \textless 0.8$,
while late-type bars are restricted to lower relative sizes,
$a_{bar} / R_{25} \lesssim 0.42$.

The distribution of absolute sizes of bars is dependent on resolution
and the galaxy population being probed.  The 2MASS sample probes the
largest, brightest and most massive galaxies in the local universe.
Figure\,6a presents our 2MASS bar sizes as a function of distance. The
median bar semi-major axis for the subset of barred spirals at a
distance $D \textless 14$\,Mpc is 1.2\,kpc.  As we include more distant
galaxies, the median bar size increases up to $\sim 3.5$\,kpc (for $D
\textless 40$ Mpc); this trend parallels the limit of bar size
detection of $\gtrsim 15 \arcsec$. The population of bars being probed
is thus affected by the bar detection limit. We must, therefore, take
into account the bar size for a proper determination of the bar
fraction.  Figure\,6b shows how the 2MASS bar fraction changes
depending on the ability to detect a particular bar size. This is
especially important for studies of bars at high redshifts as
discussed further in \S \ref{fraction_size} and \S \ref{highz}.

\subsubsection{Bar Strengths}

The higher the true ellipticity of the bar, the greater the effect of
the bar potential on the otherwise axisymmetric gravitational
potential of the disk. The maximum deprojected ellipticity thus
provides a simple assessment of the bar strength. It has been shown to
correlate well with other strength measures, such as the $Q_b$
parameter \citep{block04,laurikainen02}, which characterizes the bar
strength by measuring the maximum gravitational bar torque relative to
the galactic disk \citep{buta01}.

The typical bar in our sample has a projected ellipticity of 0.5. We
find that the distribution of ellipticities, shown in Figure\,7a, is
clearly skewed towards higher ellipticities, with a sharp drop in bar
population around $\epsilon \sim 0.75$.  By definition, our bar
detection method excludes bars with $\epsilon \lesssim 0.2$, which
explains the absence of these very weak bars. Since highly elliptical bars are the easiest to identify with our ellipse-fit method, the sharp decline in Figure\,7a is an intrinsic property of bars, indicating a lack of very strong bars in the local universe. We discuss this in more detail in \S \ref{secular}.
 
We find a weak correlation between the bar strength and the bar size,
both in terms of $a_{bar}$ and $a_{bar}/R_{25}$, as shown in Figure\,7b.
For all barred spirals in our sample we find that bar ellipticity and
relative size are correlated with a $\sim 0.99$ level of significance as
determined from the correlation r-coefficient of 0.34.  Stronger bars
appear to also be the longest ones, both in absolute size and relative
to the host disk. This trend is more significant for early-type barred
galaxies; the correlation coefficient for early-types is $r = 0.45$,
compared to $r = 0.40$ for late-types. Previous studies
\citep{martin95, laurikainen02, erwin05} found no significant
correlation between these two bar properties, though
\citet{laurikainen02} noted a slight increase in bar length with
increasing bar strength when considering only late-type galaxies.  We
find that the bar ellipticity is not correlated with the Hubble Type, the galaxy size or the $K$-band luminosity.

\section{DISCUSSION}
\subsection{Bar Fractions in the Optical and NIR}\label{fraction}

Observations in the NIR provide a better discriminant between barred
and unbarred spirals than in the optical. The RC3 classification is
based on the presence of a bar as seen in blue photographic plates:
spirals in which a bar is clearly present were classified as
strongly-barred (SB), while the bars in SAB galaxies were referred to
as ``weak bars'' or ``oval distortions'' because they are a minor distortion to the disk. In our NIR analysis, all the
intermediate SAB galaxies are classified as $barred$.  We
do not make a distinction such as the SB and the SAB classification;
galaxies are either barred or unbarred.

Our lower limit to the NIR bar fraction of 0.59 is conservative
because it is restricted to barred galaxies with both an ellipticity
and PA signature. It is consistent with the bar fraction of 59$\%$
found by LSB04, who apply a fourier decomposition method to a sample
of 158 OSUBSGS and 22 2MASS spirals with inclination $i \textless
60^\circ$.

The NIR bar fraction increases to 0.67 when we include the barred
galaxies with only the ellipticity signature.  We compare this result
to the total fraction of 0.63 (SB + SAB) in the RC3 $B$-band analysis.
Although the relative fraction of strong (SB) or weak bars (SAB)
changes based on the wavelength of observation, the overall fraction
of barred galaxies remains relatively unchanged from the $B$-band to the
$K$-band. This indicates that the bar morphology, though degraded, can
be reliably detected by eye in the $B$-band. We note that it may be
likely that fewer galaxies in the $B$-band are identified as bars by
automated algorithms due to the more irregular or patchy appearance of
galaxies in bluer bands.  Algorithms such as the ellipse-fitting
algorithm rely on a smooth light distribution and often fail for
irregular distributions.  Quantitative measurements of the change in
the bar fraction with the ellipse-fit algorithm in the $B$-band remain
to be tested.

Our results are consistent with the H-band bar fraction of 0.72
measured by E2000 in their OSUBGS sample.  In contrast to our
approach, E2000 did not make selection cuts based on inclination to
restrict their sample and, like earlier optical studies (e.g., Third
Reference Catalog of Bright Galaxies, \citealt{devau91}, hereafter,
RC3), relied on visual inspection to identify bars. W2002 report a
slightly larger H-band bar fraction of 0.79 in their sample of 72
OSUBSGS galaxies with inclination $i \textless 60^\circ$. However,
even though this method is quantitative it is based on the selection
of only two ellipses, an inner and outer one, to define a bar.  The
ellipse fitting method used here is more robust because it uses the full
two-dimensional light distribution.

Our method does not detect bars in $\sim$1/3 of our sample. There are
a number of reasons why our method may have failed to detect an
existing bar, particularly a small bar (see \S
\ref{fractionres}). However, this is unlikely to be the case in a
majority of the galaxies we classify as unbarred spirals (e.g., Figure
3e).  Models suggest that a bar may be prevented from forming if the
disk is too dynamically hot, or if the disk is not sufficiently
massive \citep{ostriker73, a05}.  Galaxies may also dissolve their
bars by accumulating a large central mass concentration or by undergoing
a merger \citep{combes85, das03, shen04}.  In either case, an unbarred
spiral should be dynamically hotter than a barred spiral. Kinematical
studies comparing unbarred and barred galaxies are needed to confirm
this hypothesis.

\begin{figure*}
\plottwo{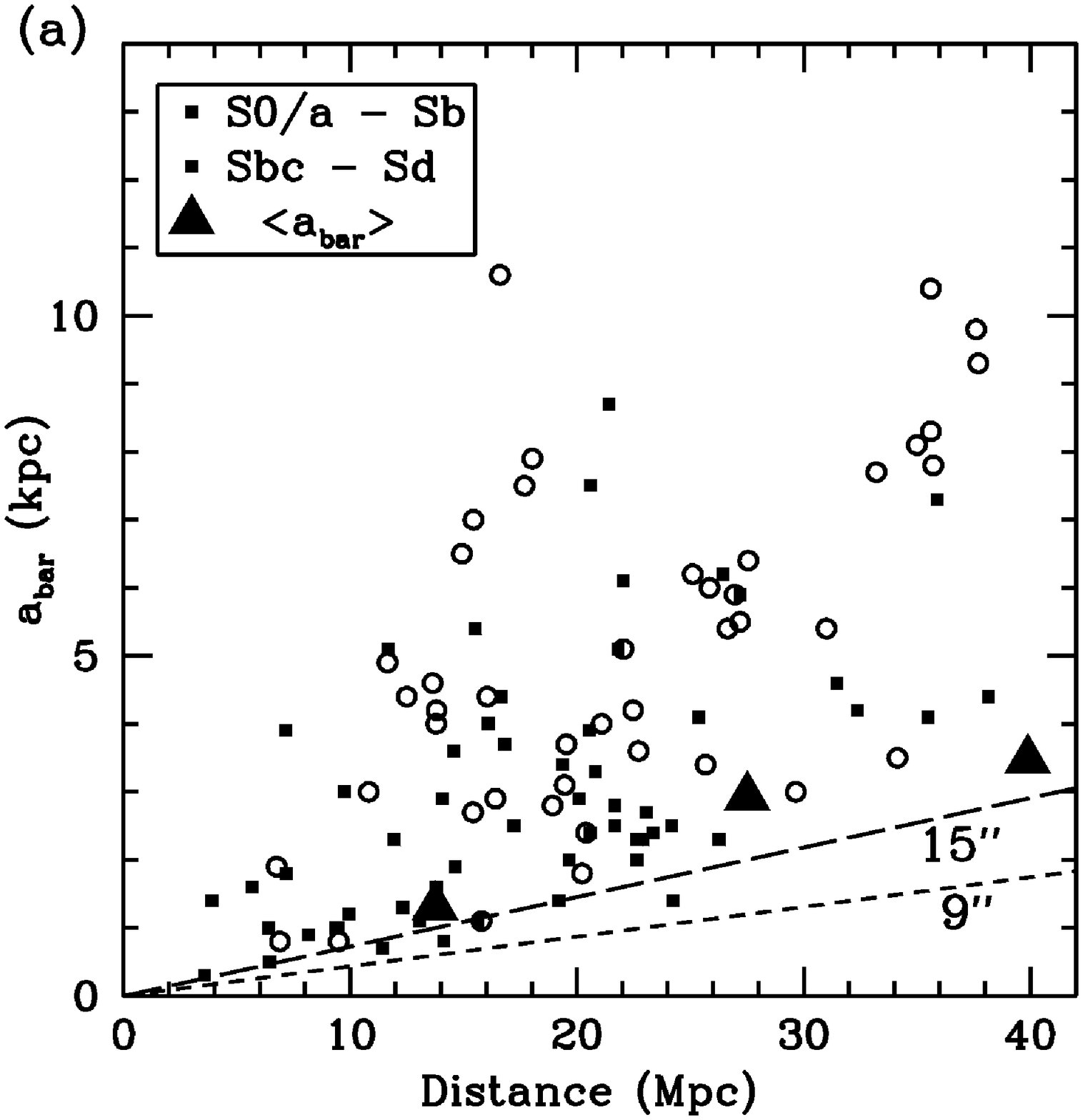}{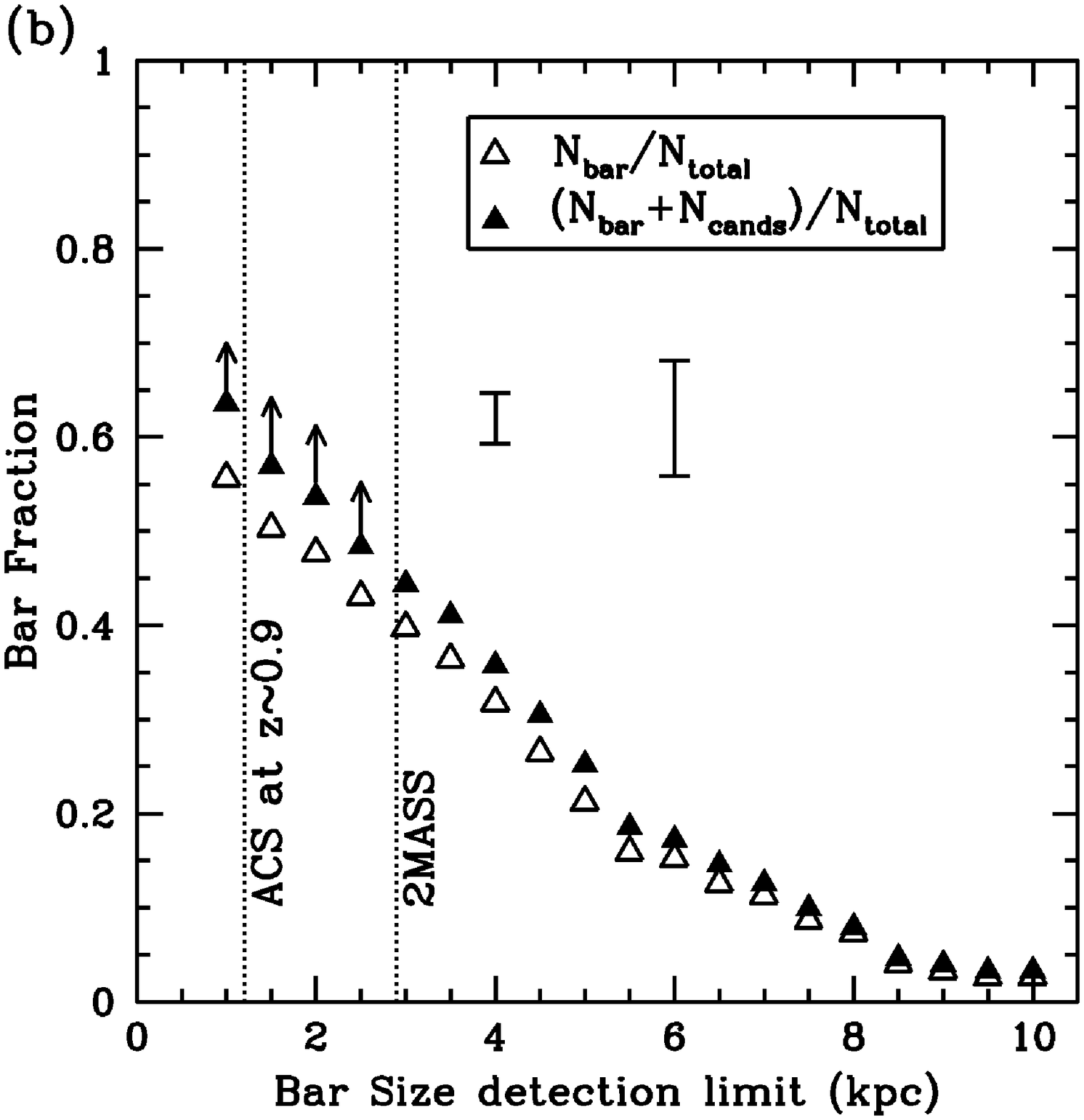}
\caption{(a) Projected bar sizes as a function of distance for our 2MASS barred sample.  Points corresponding to early-type spirals and late-type spirals are shown following the same symbol convention as Figure\,1. Sensitivity curves of our bar identification method are shown in dotted and dashed lines, corresponding to 9$\arcsec$ and 15$\arcsec$, respectively.  Cumulative median projected bar sizes are shown as large solid triangles.  The increase of this median size with distance is parallel to the size sensitivity curve of our bar detection method.  (b) Local bar fraction according to bar size detection limit.  Open triangles represent the bar fraction when we only include bars with both ellipticity and PA signatures, while solid triangles represent the bar fraction including spirals that we classified as candidates and confirmed to contain a bar by visual inspection. Using 2MASS, we are able to detect all bars with $a_{bar} > 2.9$\,kpc out to 40 Mpc.  We show this detection limit as well as the $3 \times$ PSF detection limit of ACS as dotted vertical lines. Note that the fraction changes dramatically when bar size detection limit changes. Representative Poisson errors for the bar fraction in the case of a bar size detection limit of 4\,kpc and 6\,kpc are shown, reflecting a decreasing sample size of barred galaxies with increasing bar size.}
\end{figure*}

\subsection{Secular Galaxy Evolution}\label{secular}

In the 2MASS sample, early-type barred galaxies tend to host longer
bars than late-type galaxies, both in terms of $a_{bar}$ and $a_{bar}
/ R_{25}$ (see Figure\,5), consistent with previous studies
\citep{elmegreen85,martin95,laurikainen02,erwin05}. \citet{martin95}
reports that early-type bars are on average three times larger than
late-types, slightly higher than our difference of a factor of
two. The sample of \citet{martin95} has very few early-type galaxies,
whereas our barred sample has more galaxies and a relatively
well-balanced distribution of early (44\%) and late-type spirals
(56\%).  Early-type spirals are on average more massive than
late-types and therefore may form longer bars via the bar
instability.  We also find that bars in early type galaxies have a larger
scatter in sizes than those in late-types.

Recent work by \citet{sheth05} has shown that a late-type galaxy
\textit{cannot} build an early-type bulge solely through gas inflow
along a bar; additional processes, such as mergers, are needed to
enable the transition of a late-type barred galaxy to an early-type.
One such scenario of recurrent bar formation has been proposed by
\citet{bournaud02}.  In their simulations, the second-generation bars
are shorter than their predecessors.  This may be one explanation for
the presence of short bars in the early-type galaxies in our 2MASS
sample (see Figure\,5).

The distribution of bar ellipticities shows a steep decline in bars
with high ellipticites ($\epsilon > $ 0.6) and a dearth of bars with
$\epsilon > 0.75$ (see Figure\,7a).  It is possible that these strong bars have been destroyed.  Simulations show that bars may dissolve
if a sufficiently high concentration of mass accretes at the very
centers of the disks (e.g., \citealt{norman96, shen04}).  The mass
increase is likely driven by the bar-induced gas flow
\citep{kormendy82, friedli91, friedli93, regan99}.  Since stronger
bars are more efficient in driving gas inwards, the central
concentration in these strongly barred galaxies may increase rapidly,
and may accelerate the process of bar destruction.  The decline in
strong bars in Figure\,7a may thus be an indication that stronger bars
have evolved more rapidly out of their barred state than weaker barred
spirals, and that the strongest bars have already been destroyed.

It is also possible that the drop in the high-end of the ellipticity
distribution reflects a natural limit on how thin a bar may become
either due to the presence of a bulge that limits the size of the bar
semi-minor axis, or a limit imposed by the stability of the stellar
orbits that sustain the bar. The first possibility naturally leads to
a bias of more thinner bars in late-type galaxies where bulges are
smaller than in earlier-type galaxies.  However, the ellipticity
distribution we find is even more pronounced in the late types (see
Figure\,7a). Therefore we can rule out the ``bulge hypothesis''.  The
other possibility was addressed by \citet{a83} who showed how thin
bars have more chaotic orbits. This may also explain the lack of thin
bars beyond a certain ellipticity at which a bar becomes unable to
sustain itself.

\begin{figure*}
\plottwo{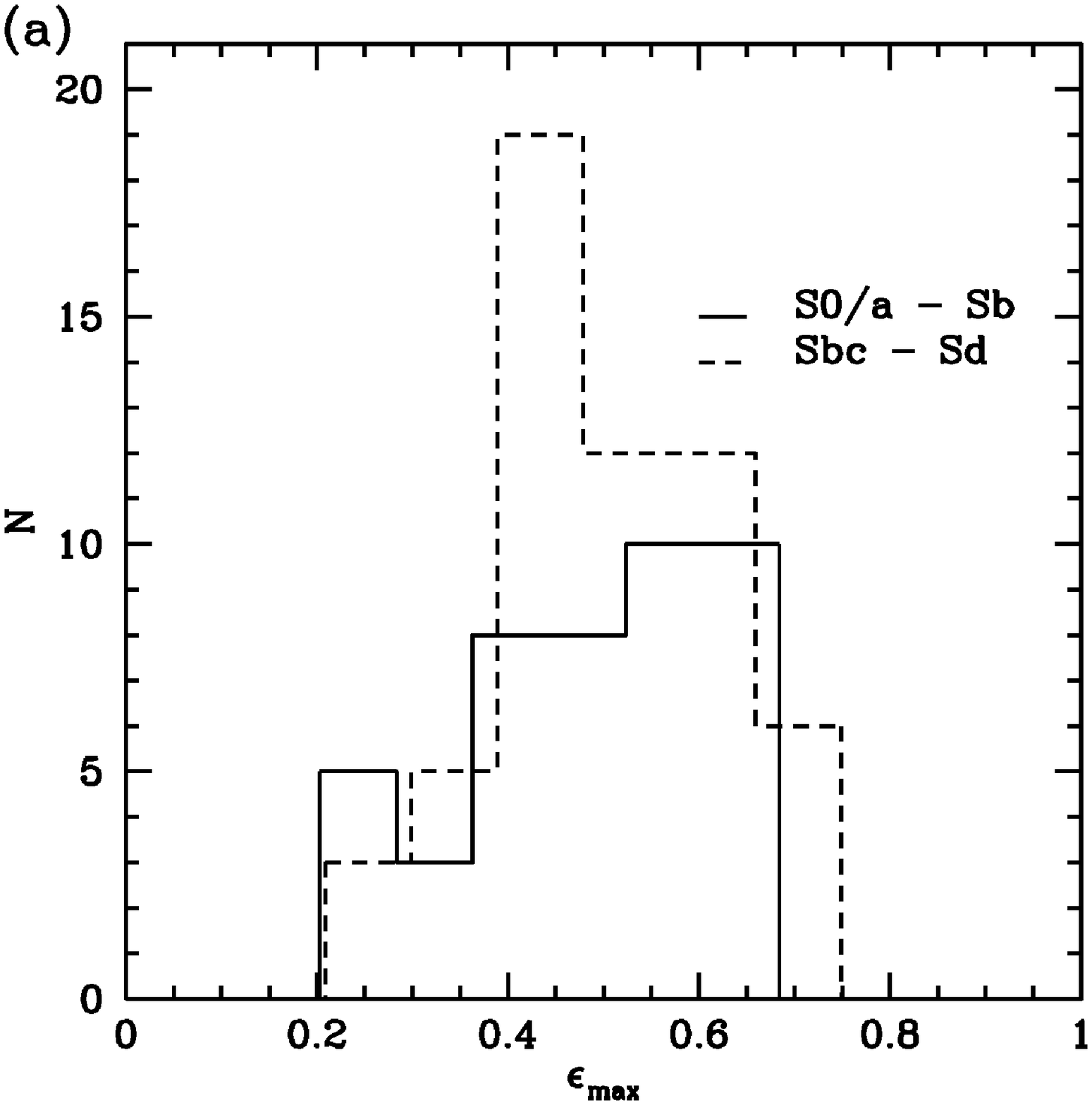}{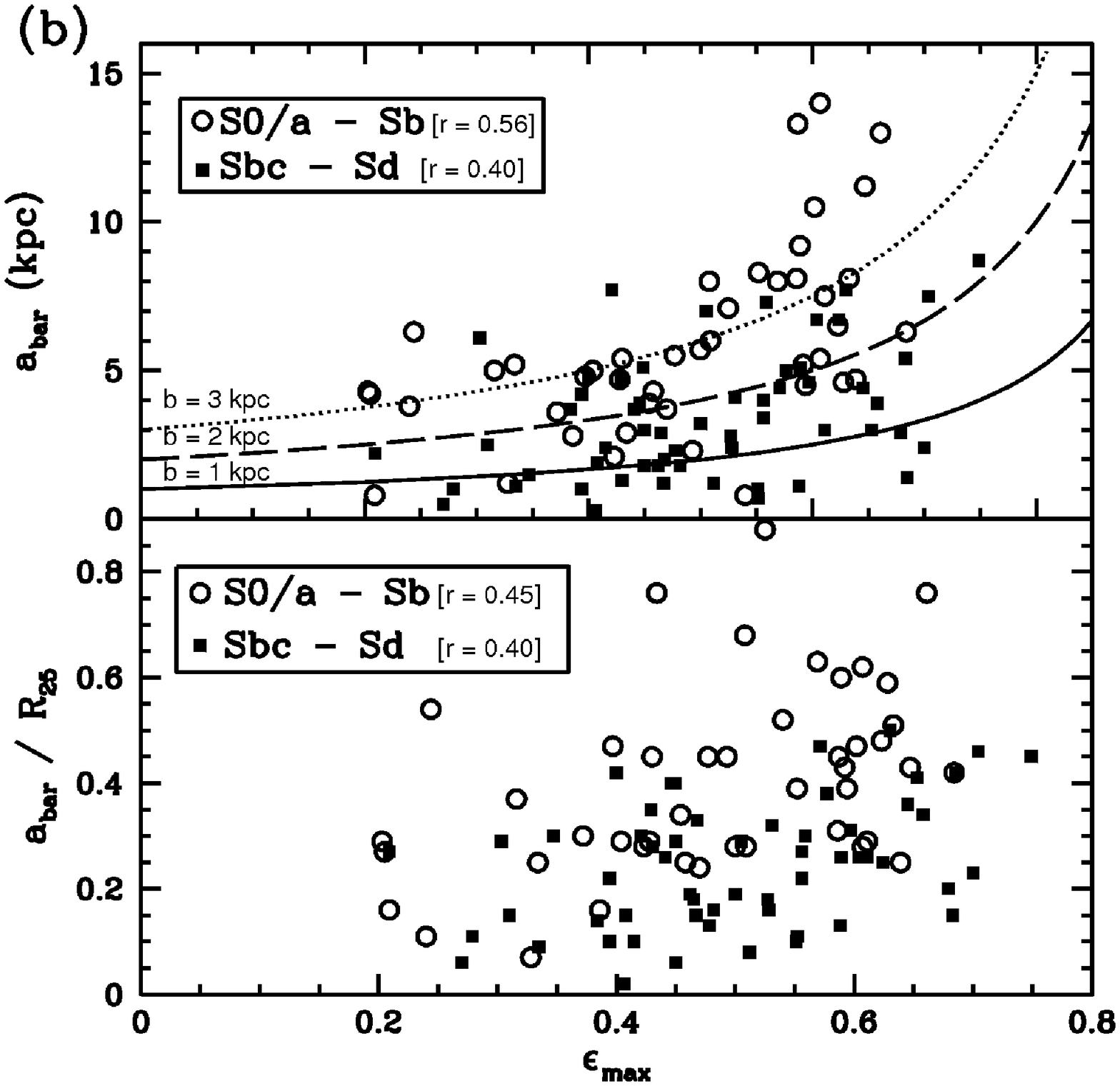}
\caption{(a) Distribution of projected ellipticies for 2MASS bars.    The distribution for early-type galaxies is shown as a solid line and that of late-types, as a dashed line.  Both histograms show a smooth increase in number for increasing ellipticity and a sharp decline for bars with $\epsilon$ $\gtrsim$ 0.75.  Highly elliptical bars correspond to thin, strong bars. (b) Top: $a_{bar}$ as a function of projected ellipticity.  Symbols for early-type and late-type bars follow the same convention as Figure\,1. We find a weak trend between $a_{bar}$ and $\epsilon_{max}$, which shows that larger bars are stronger; the trend is more evident in the early type galaxies. Curves of constant semi-minor axis values are shown in solid, dashed and dotted lines for b = 1, 2 and 3\,kpc, respectively. Note that at a given ellipticity the early type galaxies lie above the curves. Bottom: A similar plot for $a_{bar} / R_{25}$.}
\end{figure*}

\subsection{Bar Properties}

We find a weak correlation between the bar size and the bar
ellipticities (see Figure\,7b) with longer bars having somewhat greater
ellipticities.  This trend is likely to be stronger because the
measured ellipticities are underestimated for big bulged galaxies with
thin bars, e.g., bars whose semi-minor axis is smaller than the bulge
radius.  This is because the ellipse fitting routine does not
decompose the bulge and disk, and cannot measure an ellipticity with a
semi-minor axis smaller than the bulge radius. Therefore, the true
ellipticity of the thinner early type spirals is likely to be even
higher which strengthens the correlation shown in Figure\,7b.  The expected correlation between ellipticity and the semi-major axis for a constant semi-minor axis is shown by the curves in Figure\,7b.  The measured ellipticity for early type spirals are consistently above these curves.This indicates that the bar semi-major axis ($a_{bar}$) changes
more rapidly with $\epsilon_{bar}$ than the semi-minor axis
($b_{bar}$).  

\subsection{The Change in Bar Fraction As a Function of Bar Size}\label{fraction_size}

In deriving a bar fraction it is critical to define the bar population
that is being studied.  At a fixed angular resolution, the most
important selection effect is the decreasing ability to detect smaller
bars with increasing distance. This is a serious problem for high
redshift studies aimed at measuring the evolution of the bar fraction,
as shown in Figure\,3 of \citet{sheth03}.  With our 2MASS sample we can
estimate how the bar fraction must change with changing spatial
resolutions.  In Figure\,6b we show the cumulative bar fraction for
different detection limits of the bar size. The fraction of barred
spirals drops from 67\% over the entire sample to 36\% when we only
count bars larger than 4\,kpc, and a mere 8\% for bars larger than 8
kpc. An important caveat to be noted is that even with our large NIR
data set, we are incomplete for the small bars.  Our bar size
detection limit changes from 0.5\,kpc to 3\,kpc over the volume surveyed
(D $ \textless $ 40 Mpc). Hence we are only complete for bars with
$a_{bar} \gtrsim 3$\,kpc, but this is more than adequate for comparison
to high redshift observations which typically have a similar spatial
resolution.  

The 2MASS sample thus offers a detailed local comparison sample for
the high redshift studies.  As noted in the previous section, the
effect of bandshifting from the $K$-band to the $B$-band is likely to be
minimal and thus 2MASS is a local anchor for the studies of galaxy
evolution in the rest frame optical.

\subsection{The Local Bar Fraction as an anchor for high redshift studies}\label{highz}

Our NIR results show that there is not a significant population of
veiled barred spirals in the optical, as we discussed in \S
\ref{fraction}.  In other words, the HST $I$- and $z$-band data now
available for high redshift studies can be used to trace the bar
fraction to at least a redshift of $\sim 0.7-0.8$ where the observed
$I$-band data images galaxies in the rest frame $B$-band. We strongly caution
against interpretation of bar fractions for redshifts higher than $z
\sim 0.8$ because there is strong evidence that bars completely
disappear in the ultra-violet (UV); an example is shown in Figure\,1 of
\citet{sheth03}.  Other examples are evident in the GALEX images,
which show no bars in nearby galaxies (Gil de Paz, private
communication)\footnote{We have looked at publicly available GALEX
  images of a subset of the 2MASS spiral sample and have found no
  indication of bars.}. Since UV light is dominated by young ($
\textless 10^8$ yrs) star forming regions, it may not trace the
gravitational potential and bar signatures are likely to be absent in
the UV.  Any measurement of the bar fraction at $z > 0.8$ in the
rest-frame UV is therefore likely to suffer from an abrupt decline in
the number of bars as they disappear shortwards of the Balmer break.
Additionally, at $z > 0.8$, observations are affected by coarse linear
resolution (see Figure\,3 in \citealt{sheth03}) and the steep decline
in the cosmological dimming of the surface brightness.  We note that
this study, however, is a good analog for high redshift studies
because the angular resolution and shallow depth of 2MASS are similar
to the limitations of high redshift data described above.

Various authors have studied the evolution of the bar fraction with
redshift.  \citet{sheth03} first challenged the previous findings that
barred spirals were completely absent at high redshifts
\citep{abraham99, vandenBergh96, vandenBergh00, vandenBergh01}.
\citet{sheth03} studied all galaxies in the HDF-N in the WFPC2 $V$, $I$
and NICMOS-$H$ bands to investigate whether bars were more apparent at
the longer NICMOS wavelengths.  At $z > 0.7$, the coarse NICMOS data
are only able to probe the biggest spirals.  Here they find 3/31
barred spirals with an average bar semi-major axis of 6.4\,kpc.  This
fraction is comparable to the 2MASS fraction of 15\% for bars with
$a_{bar} > 6.5$\,kpc, but we note that both samples suffer from small
number statistics - large bars are rare and require very large search
volumes. With these caveats, \citet{sheth03} concluded that there was no strong
evidence for an absence of long bars at high redshift.

Following \citet{sheth03}, two studies have examined the fraction of
bars with higher resolution optical ACS data.  \citet{elmegreen04}
studied 186 galaxies from the Tadpole field out to $z \sim 1.2$ and
reported a nearly constant bar fraction of 0.23 out to $z \sim
1.1$. \citet{jogee04} analysed 258 galaxies from the GEMS survey and
reported a similar constant bar fraction of 0.30 out to $z \sim
1$. These studies argue that at high redshifts the bar fraction is
measured in the rest-frame $B$-band and is therefore comparable to
the 0.35 value measured for SB galaxies in the RC3.  However they also claim that
the fraction is the same at intermediate redshifts.  At redshifts of
$0.2 \textless z \textless 0.7$, the ACS angular resolution and data
quality are superb and selection effects noted by \citet{elmegreen04}
and \citet{jogee04} (e.g. surface brightness dimming, coarse
resolution, poor signal to noise) ought to be minimal. Moreover, the
rest-frame wavelength data for galaxies at intermediate redshifts is
longwards of the $B$-band.  Therefore we would have expected these
studies to detect both the SB spirals and the SAB spirals.  Their bar fraction should have been comparable to our local NIR result. However, for their bar
detection threshold of 1.2\,kpc, their measured bar fraction is lower
than the 2MASS value of 0.58 by almost a factor of two.

If the bar fraction is indeed a constant out to $z \sim 0.8$, as
suggested by \citet{elmegreen04} and \citet{jogee04}, we would expect
the fraction of bars at $z \sim 0.7-0.8$, to be the same as the local
(SB + SAB) RC3 fraction measured by eye.  One possibility for this
discrepancy is the small sample sizes of these existing studies.  All
of them have less than 300 galaxies over the entire redshift range
with which they have attempted to constrain the evolution of the bar
fraction.  Another possibility is that the bars and galaxies are
evolving.  For example, if galaxies at $z \sim 1$ are smaller than
local galaxies as suggested by \citet{elmegreen05}, then the bars at
these redshifts may also be smaller, and may be undetected by these
studies.  However, the difference between the local bar fraction and the fraction measured at these higher redshifts may also indicate a potential evolution of the bar fraction with redshift. Thus the perceived non-evolution of the bar fraction may be
a misinterpretation of the active evolution in the bar fraction. In
order to explore this possibility, we are analyzing the 2-square
degree COSMOS field to study in detail the evolution of bar fraction
to $z \sim 0.8$.  The COSMOS data are extremely deep (only half a
magnitude shallower than GOODS $z$-band data), and have over 4000
L$^{*}$ galaxies at $z <  0.8$. With this large and deep dataset we will
quantify the evolution of the bar fraction and bar sizes while using
the results presented here as a local anchor and guide for
understanding galaxy evolution.

\section{CONCLUSIONS}

We have performed a detailed study of 151 2MASS spirals out to 40 Mpc
using an ellipse-fitting technique to derive the local fraction of
barred spirals and to characterize their properties.  We discussed in
detail the advantages and shortcomings of our technique for its use as
a practical approach in analyzing large data sets for which visual
inspection of individual images becomes rapidly inefficient.  We
discussed how our detailed analysis of the local bar population sets the
groundwork for studies of galaxy evolution at higher redshifts.

Our main results are:

1. We have found a lower limit to the NIR fraction of barred galaxies
of 0.59 in the local universe. By complementing our automated bar
detection method with visual inspection, the total bar fraction
increases to 0.67. This suggests that the bar fraction in the NIR is
not significantly different than the bar fraction of 0.63 (SB + SAB)
in the optical. This is a promising result for work on the available, expansive
datasets of high redshift galaxies that probe rest-frame
optical wavelengths.

2.  The typical 2MASS bar extends out to $a_{bar} / R_{25}\,\sim\,0.3$
in radius relative to the underlying blue disk.

3.  2MASS bars have a median semi-major axis of 4.2\,kpc. The 2MASS
coarse angular resolution and shallow depth favors the detection of
larger bars.
    
4.  A weak trend relating bar strength and size appears to be present.
The correlation between the bar ellipticity, $\epsilon_{max}$, and $a_{bar}/R_{25}$ is
stronger ($r = 0.46$ ) for early-type galaxies and shows that
$a_{bar}$ (and $a_{bar}/R_{25}$) evolves more rapidly with increasing
ellipticity than the bar semi-minor axis, $b_{bar}$, whose size is regulated by the bulge radius.

5.  We find that the mean $a_{bar}/R_{25}$ in early-type spirals is
two times larger than in late-types, confirming that bars in
early-type galaxies are larger than in late-types.

6. We show how the measured bar fraction depends critically on the
population of bar sizes that can be probed by the observations and
detection method. Whereas the bar fraction is 0.64 for bars longer
than $a_{bar} = 1$\,kpc, the fraction decreases to 0.44 for $a_{bar}\,>\,3$\,kpc.  Careful consideration of this bar detection threshold must be
taken into account for high redshift studies of the bar fraction.

7.  Even after taking into account the ability to resolve bars, the
locally measured bar fraction is higher by nearly a factor of two than
the bar fraction at $z \sim 1$ reported by recent optical ACS
studies. Our results do not support the conclusion in these studies
that the bar fraction remains constant out to $z \sim 1$. The
difference in the bar fraction with these high-redshift studies may be
due to small number statistics of the these studies (all have $< 300$
galaxies from $0 < z < 1$).  It may, however, indicate a potential
evolution of the bar fraction with redshift.  We are investigating the
evolution of the bar fraction with over four thousand L$^{*}$ spirals
from the COSMOS data set.

\acknowledgements

We thank our anonymous referee for useful comments that have greatly improved this paper. We are grateful to Michael Regan and Peter Teuben for our discussions
on the analysis of this study. We thank Bruce Elmegreen, Linda Sparke,
Ron Buta, Leslie Hunt, Peter Erwin, Shardha Jogee, Paul Martini and
Luis Ho for their helpful insights.  We would also like thank David
Block for organizing an excellent conference, ``Penetrating Bars Through Masks of
Cosmic Dust'', and for his hospitality.

This publication makes use of data products from the Two Micron All
Sky Survey, which is a joint project of the University of
Massachusetts and the Infrared Processing and Analysis
Center/California Institute of Technology, funded by the National
Aeronautics and Space Administration and the National Science
Foundation.

\begin{deluxetable*}{rcccccccc}
\tabletypesize{\scriptsize}
\tablecaption{2MASS Spiral Sample\label{sampletab}}
\tablehead{
  \colhead{Galaxy} & \colhead{$\alpha$(J2000)\tablenotemark{a}}  &  \colhead{$\delta$(J2000)\tablenotemark{a}}  &  \colhead{$K_{abs}$} & \colhead{$K$\tablenotemark{b}} &  \colhead{D\tablenotemark{c}} &  \colhead{$R_{25}$\tablenotemark{a}}  &  \colhead{$i$\tablenotemark{a}}  &  \colhead{PA\tablenotemark{a}}     \\
\colhead{}  &  \colhead{(hours)}  &  \colhead{($\deg$)} &  \colhead{(mag)} &  \colhead{(mag)}  &  \colhead{(Mpc)} &  \colhead{(kpc)}  &  \colhead{($\deg$)}  &  \colhead{($\deg$)}}
\startdata
\tableline
 &  &  &  &  S0/a Spirals &  &  &  &   \\
NGC1291 & 3.289 & -41.108 & -24.67 & 5.66 & 11.6 & 17.12 & 33 & 156  \\
NGC1317 & 3.379 & -37.103 & -24.39 & 7.74 & 26.6 & 11.68 & 32 & 72  \\
NGC1326 & 3.399 & -36.464 & -23.93 & 7.45 & 18.9 & 12.04 & 49 & 67  \\
NGC2217 & 6.361 & -27.234 & -24.67 & 7.09 & 22.5 & 16.19 & 28 & 0\tablenotemark{d}  \\
NGC2681 & 8.892 & 51.313 & -22.45 & 7.43 & 9.5 & 5.02 & 0 & 3\tablenotemark{e}  \\
NGC2655 & 8.927 & 78.224 & -24.49 & 6.95 & 19.5 & 13.63 & 54 & 16  \\
NGC5101 & 13.363 & -27.430 & -24.91 & 7.16 & 25.8 & 21.83 & 40 & 124  \\
 &  &  &  &  Sa Spirals &  &  &  &   \\
NGC1022 & 2.642 & -6.677 & -23.09 & 8.44 & 20.2 & 7.53 & 46 & 68 \\ 
NGC1367 & 3.584 & -24.934 & -23.92 & 7.63 & 20.4 & 17.59 & 51 & 134 \\ 
NGC3169 & 10.237 & 3.466 & -23.88 & 7.28 & 17.1 & 11.57 & 63 & 45 \\
NGC3718 & 11.543 & 53.068 & -22.94 & 7.76 & 13.8 & 15.15 & 63 & 180 \\
NGC4314 & 12.376 & 29.895 & -23.22 & 7.45 & 13.6 & 7.79 & 19 & 155\tablenotemark{e} \\
NGC4491 & 12.516 & 11.484 & -19.31 & 9.88 & 6.9 & 1.61 & 64 & 147 \\
NGC5728 & 14.707 & -17.253 & -24.77 & 8.17 & 37.6 & 17.25 & 58 & 180 \\
NGC7172 & 22.034 & -31.870 & -24.45 & 8.32 & 35.7 & 11.76 & 58 & 105 \\
NGC7727 & 23.665 & -12.292 & -24.36 & 7.69 & 25.7 & 15.38 & 40 & 28\tablenotemark{e} \\
 &  &  &  &  Sab Spirals &  &  &  &  \\
NGC1350 & 3.519 & -33.629 & -24.71 & 7.40 & 25.1 & 21.45 & 64 & 11 \\
NGC1398 & 3.648 & -26.338 & -24.96 & 6.50 & 19.5 & 21.26 & 45 & 100 \\
NGC1433 & 3.700 & -47.222 & -23.81 & 7.06 & 14.9 & 13.47 & 48 & 95 \\
NGC1512 & 4.065 & -43.349 & -23.00 & 7.49 & 12.5 & 18.55 & 61 & 56 \\
NGC2146 & 6.310 & 78.356 & -23.40 & 7.06 & 12.4 & 9.91 & 58 & 123 \\
NGC2566 & 8.313 & -25.499 & -24.02 & 7.77 & 22.8 & 12.73 & 60 & 69 \\
NGC2775 & 9.172 & 7.038 & -24.31 & 7.06 & 18.8 & 12.21 & 40 & 160 \\
NGC2985 & 9.840 & 72.280 & -23.96 & 7.36 & 18.3 & 11.07 & 39 & 176 \\
NGC3049 & 9.914 & 9.272 & -21.63 & 9.96 & 20.8 & 6.51 & 56 & 27 \\
NGC3031 (M81) & 9.926 & 69.066 & -24.84 & 3.83 & 5.4 & 17.83 & 63 & 157 \\
NGC3368 (M96) & 10.779 & 11.820 & -24.16 & 6.32 & 12.5 & 14.04 & 51 & 5 \\
NGC3898 & 11.821 & 56.084 & -23.41 & 7.66 & 16.3 & 8.76 & 53 & 108 \\
NGC4151 & 12.176 & 39.406 & -23.32 & 7.38 & 13.8 & 12.15 & 40 & 50 \\
NGC4448 & 12.471 & 28.621 & -21.98 & 7.81 & 8.9 & 3.29 & 60 & 98 \\
NGC4450 & 12.475 & 17.085 & -25.12 & 7.05 & 27.2 & 20.19 & 43 & 173 \\
NGC4750 & 12.835 & 72.875 & -23.74 & 8.02 & 22.5 & 7.28 & 31 & 142 \\
NGC4725 & 12.841 & 25.501 & -24.95 & 6.17 & 16.6 & 26.23 & 51 & 33 \\
NGC4736 (M94) & 12.848 & 41.120 & -24.66 & 5.11 & 9.0 & 16.55 & 30 & 105 \\
NGC4941 & 13.070 & -5.552 & -22.73 & 8.22 & 15.4 & 8.02 & 54 & 23 \\
NGC5317 & 13.888 & 33.491 & -24.80 & 7.80 & 16.4 & 3.83 & 43 & 58 \\
NGC7217 & 22.131 & 31.359 & -23.75 & 6.85 & 13.2 & 7.01 & 34 & 88 \\
NGC7552 & 23.270 & -42.584 & -24.18 & 7.54 & 22.1 & 11.94 & 51 & 103 \\
 &  &  &  &  Sb Spirals &  &  &  &  \\
NGC210 & 0.676 & -13.874 & -23.39 & 8.39 & 22.7 & 15.63 & 54 & 163 \\
NGC488 & 1.363 & 5.257 & -25.53 & 6.96 & 31.5 & 24.63 & 43 & 15 \\
NGC772 & 1.989 & 19.008 & -25.47 & 7.20 & 34.1 & 35.23 & 59 & 131 \\
NGC986 & 2.559 & -39.045 & -24.42 & 7.78 & 27.6 & 15.84 & 57 & 28 \\
NGC1068 & 2.711 & -0.013 & -25.20 & 5.79 & 15.8 & 17.22 & 29 & 73 \\
NGC1097 & 2.772 & -30.275 & -24.98 & 6.25 & 17.7 & 25.59 & 51 & 140 \\
NGC1365 & 3.560 & -36.141 & -25.43 & 6.37 & 18.0 & 29.56 & 55 & 12 \\
NGC1530 & 4.391 & 75.296 & -24.39 & 8.29 & 37.7 & 22.70 & 59 & 81 \\
NGC2090 & 5.784 & -34.251 & -22.50 & 8.05 & 12.9 & 11.31 & 64 & 17 \\
NGC3227 & 10.392 & 19.865 & -23.39 & 7.64 & 16.0 & 11.56 & 59 & 157 \\
NGC3351 & 10.733 & 11.704 & -23.50 & 6.67 & 10.8 & 11.95 & 57 & 13 \\
NGC3583 & 11.236 & 48.319 & -23.98 & 8.38 & 29.7 & 10.23 & 55 & 134 \\
NGC3673 & 11.420 & -26.737 & -23.65 & 8.51 & 27.0 & 15.04 & 59 & 70 \\
NGC3675 & 11.436 & 43.586 & -23.28 & 6.86 & 10.7 & 9.67 & 58 & 178 \\
NGC4102 & 12.107 & 52.711 & -22.63 & 7.72 & 11.7 & 5.30 & 59 & 38 \\
NGC4548 (M88) & 12.533 & 14.420 & -26.24 & 6.27 & 31.7 & 31.28 & 60 & 138 \\
NGC4548 (M91) & 12.591 & 14.496 & -22.03 & 7.12 & 6.8 & 5.11 & 35 & 150 \\
NGC4579 & 12.629 & 11.818 & -25.13 & 6.49 & 21.1 & 17.04 & 39 & 89 \\
NGC4593 & 12.661 & -5.344 & -24.71 & 7.99 & 35.6 & 15.42 & 55 & 38 \\
NGC4595 & 12.664 & 15.298 & -19.69 & 10.03 & 7.7 & 1.87 & 52 & 110 \\
NGC5383 & 13.951 & 41.846 & -23.93 & 8.54 & 33.2 & 12.79 & 38 & 107 \\
NGC5850 & 15.119 & 1.545 & -24.65 & 8.10 & 35.6 & 22.24 & 41 & 107 \\
NGC5985 & 15.660 & 59.332 & -24.57 & 8.15 & 35.0 & 26.16 & 64 & 15 \\
NGC6300 & 17.283 & -62.820 & -24.01 & 6.93 & 15.4 & 11.29 & 54 & 119 \\
NGC6902 & 20.408 & -43.654 & -24.33 & 8.61 & 38.7 & 29.01 & 43 & 158 \\
NGC7606 & 23.318 & -8.485 & -24.82 & 7.64 & 31.0 & 19.32 & 60 & 145 \\
\enddata
\end{deluxetable*}

\setcounter{table}{0}
\begin{deluxetable*}{rcccccccc}
\tabletypesize{\scriptsize}
\tablecaption{2MASS Spiral Sample $-$ Continued}
\tablehead{
  \colhead{Galaxy} & \colhead{$\alpha$(J2000)\tablenotemark{a}}  &  \colhead{$\delta$(J2000)\tablenotemark{a}}  &  \colhead{$K_{abs}$} & \colhead{$K$\tablenotemark{b}} &  \colhead{D\tablenotemark{c}} &  \colhead{$R_{25}$\tablenotemark{a}}  &  \colhead{$i$\tablenotemark{a}}  &  \colhead{PA\tablenotemark{a}}     \\
\colhead{}  &  \colhead{(hours)}  &  \colhead{($\deg$)} &  \colhead{(mag)} &  \colhead{(mag)}  &  \colhead{(Mpc)} &  \colhead{(kpc)}  &  \colhead{($\deg$)}  &  \colhead{($\deg$)}}

\startdata
\tableline
 &  &  &  &  Sbc Spirals &  &  &  &  \\
NGC289 & 0.878 & -31.206 & -23.78 & 8.00 & 22.6 & 18.05 & 40 & 129 \\
NGC613 & 1.572 & -29.418 & -24.54 & 7.03 & 20.6 & 16.28 & 50 & 119 \\
NGC1300 & 3.328 & -19.411 & -24.14 & 7.56 & 21.4 & 19.55 & 59 & 101 \\
NGC1566 & 4.333 & -54.938 & -24.71 & 6.89 & 20.8 & 26.42 & 44 & 32 \\
NGC1672 & 4.762 & -59.248 & -24.34 & 7.02 & 15.5 & 15.42 & 46 & 170\tablenotemark{d} \\
NGC2207 & 6.273 & -21.373 & -24.72 & 8.19 & 38.2 & 25.42 & 61 & 116 \\
NGC2336 & 7.451 & 80.178 & -24.72 & 7.70 & 30.6 & 28.30 & 57 & 178 \\
NGC2442 & 7.607 & -69.531 & -24.65 & 6.87 & 20.1 & 15.72 & 50 & 23 \\
NGC2559 & 8.285 & -27.457 & -24.26 & 7.42 & 21.7 & 12.95 & 61 & 4 \\
NGC3344 & 10.725 & 24.922 & -22.12 & 7.44 & 8.1 & 8.51 & 17 & 140\tablenotemark{d} \\
NGC3521 & 11.097 & -0.035 & -24.46 & 5.78 & 11.2 & 17.63 & 63 & 164 \\
NGC3642 & 11.372 & 59.074 & -22.74 & 8.97 & 22.0 & 17.73 & 35 & 105 \\
NGC3882 & 11.768 & -56.391 & -23.71 & 8.31 & 25.4 & 9.40 & 55 & 107 \\
NGC3953 & 11.897 & 52.327 & -23.78 & 7.05 & 14.6 & 16.29 & 63 & 13 \\
NGC3992 (M109) & 11.960 & 53.375 & -23.88 & 6.94 & 14.6 & 14.81 & 52 & 78 \\
NGC4051 & 12.053 & 44.531 & -22.27 & 7.67 & 9.7 & 7.23 & 29 & 135 \\
NGC4303 (M61) & 12.365 & 4.475 & -24.85 & 6.84 & 21.8 & 19.65 & 19 & 162 \\
NGC4321 (M100) & 12.382 & 15.823 & -25.13 & 6.59 & 22.0 & 24.25 & 38 & 155\tablenotemark{d} \\
NGC4567 (VV219B) & 12.609 & 11.258 & -24.19 & 8.30 & 31.5 & 12.58 & 43 \\
NGC4602 & 12.677 & -5.133 & -24.20 & 8.54 & 35.3 & 11.24 & 63 &  \\
NGC4639 & 12.715 & 13.257 & -21.95 & 8.75 & 13.8 & 6.05 & 52 & 134 \\
NGC4930 & 13.068 & -41.411 & -23.94 & 8.83 & 35.9 & 24.19 & 46 & 44 \\
NGC5055 (M63) & 13.264 & 42.029 & -23.61 & 5.61 & 7.0 & 13.47 & 55 & 102 \\
NGC5054 & 13.283 & -16.634 & -24.33 & 7.59 & 24.2 & 16.87 & 54 & 171 \\
NGC5194 (M51A) & 13.498 & 47.195 & -23.54 & 5.50 & 6.4 & 9.22 & 47 & 163 \\
NGC5248 & 13.626 & 8.885 & -23.77 & 7.25 & 16.0 & 13.24 & 50 & 125 \\
NGC5247 & 13.634 & -17.884 & -23.85 & 7.53 & 18.8 & 14.62 & 43 & 170 \\
NGC5371 & 13.928 & 40.462 & -25.14 & 7.61 & 35.5 & 21.36 & 48 & 2 \\
NGC5347 & 13.937 & 5.015 & -21.54 & 9.65 & 17.2 & 15.06 & 57 & 121 \\
NGC5713 & 14.670 & -0.290 & -23.77 & 8.33 & 26.3 & 10.42 & 40 & 12 \\
NGC5921 & 15.366 & 5.071 & -23.47 & 8.10 & 20.5 & 13.02 & 44 & 140 \\
NGC6221 & 16.880 & -59.216 & -24.45 & 7.12 & 20.6 & 13.27 & 62 & 1 \\
NGC6384 & 17.540 & 7.061 & -24.29 & 7.53 & 23.1 & 16.65 & 60 & 14 \\
NGC6744 & 19.163 & -63.858 & -24.40 & 5.94 & 11.7 & 35.54 & 49 & 16 \\
 &  &  &  &  Sc Spirals &  &  &  &  \\
NGC578 & 1.508 & -22.667 & -23.19 & 8.59 & 22.6 & 15.30 & 54 & 115 \\
NGC628 & 1.612 & 15.783 & -22.95 & 6.85 & 9.1 & 14.01 & 23 & 70\tablenotemark{e} \\
NGC864 & 2.258 & 6.002 & -23.15 & 8.53 & 21.7 & 14.27 & 48 & 24 \\
NGC1073 & 2.728 & 1.376 & -22.15 & 8.98 & 16.8 & 11.40 & 38 & 32 \\
NGC1084 & 2.767 & -7.579 & -23.52 & 7.93 & 19.5 & 8.94 & 52 & 35 \\
NGC1187 & 3.044 & -22.868 & -23.33 & 8.10 & 19.4 & 15.44 & 48 & 129 \\
NGC1232 & 3.163 & -20.581 & -24.46 & 7.38 & 23.4 & 24.44 & 30 & 99 \\
NGC1637 & 4.691 & -2.857 & -22.01 & 7.97 & 9.9 & 4.71 & 32 & 13 \\
NGC1809 & 5.035 & -69.567 & -21.95 & 9.35 & 18.2 & 3.05 & 51 & 143 \\
NGC2835 & 9.298 & -22.356 & -22.54 & 7.92 & 12.3 & 12.00 & 50 & 3 \\
NGC2997 & 9.761 & -31.191 & -24.48 & 6.41 & 15.1 & 21.56 & 46 & 97 \\
NGC2976 & 9.788 & 67.917 & -20.43 & 7.52 & 3.9 & 3.28 & 61 & 143 \\
NGC3338 & 10.702 & 13.747 & -23.16 & 8.13 & 18.1 & 10.74 & 55 & 100\tablenotemark{f} \\
NGC3359 & 10.777 & 63.224 & -22.12 & 8.62 & 14.1 & 14.96 & 52 & 176 \\
NGC3486 & 11.007 & 28.975 & -21.88 & 8.00 & 9.5 & 9.29 & 45 & 80 \\
NGC3614 & 11.306 & 45.748 & -22.92 & 9.63 & 32.4 & 19.40 & 50 & 87 \\
NGC3631 & 11.351 & 53.169 & -23.05 & 7.99 & 16.1 & 11.44 & 35 & 118 \\
NGC3726 & 11.556 & 47.029 & -22.60 & 7.78 & 11.9 & 10.49 & 49 & 14 \\
NGC3938 & 11.880 & 44.122 & -22.44 & 7.81 & 11.2 & 8.11 & 13 &  \\
NGC4254 & 12.314 & 14.417 & -25.70 & 6.93 & 33.5 & 26.05 & 29 &  \\
NGC4535 & 12.572 & 8.198 & -24.79 & 7.38 & 27.2 & 27.30 & 41 & 180 \\
NGC4647 & 12.726 & 11.583 & -23.42 & 8.05 & 19.7 & 8.02 & 34 & 125 \\
NGC5236 (M83) & 13.617 & -29.865 & -24.65 & 4.62 & 7.1 & 14.47 & 24 & 48\tablenotemark{e} \\
NGC5426 & 14.057 & -6.067 & -23.31 & 9.50 & 36.4 & 15.80 & 62 & 1 \\
NGC5427 & 14.057 & -6.031 & -24.21 & 8.59 & 36.4 & 15.19 & 42 &  \\
NGC5643 & 14.545 & -44.174 & -23.94 & 7.17 & 16.7 & 11.97 & 30 & 88 \\
NGC6215 & 16.852 & -58.993 & -23.40 & 8.28 & 21.6 & 6.67 & 44 & 83 \\
\enddata
\end{deluxetable*}

\setcounter{table}{0}
\begin{deluxetable*}{rcccccccc}
\tabletypesize{\scriptsize}
\tablecaption{2MASS Spiral Sample $-$ Continued}
\tablehead{
  \colhead{Galaxy} & \colhead{$\alpha$(J2000)\tablenotemark{a}}  &  \colhead{$\delta$(J2000)\tablenotemark{a}}  &  \colhead{$K_{abs}$} & \colhead{$K$\tablenotemark{b}} &  \colhead{D\tablenotemark{c}} &  \colhead{$R_{25}$\tablenotemark{a}}  &  \colhead{$i$\tablenotemark{a}}  &  \colhead{PA\tablenotemark{a}}     \\
\colhead{}  &  \colhead{(hours)}  &  \colhead{($\deg$)} &  \colhead{(mag)} &  \colhead{(mag)}  &  \colhead{(Mpc)} &  \colhead{(kpc)}  &  \colhead{($\deg$)}  &  \colhead{($\deg$)}}

\startdata
\tableline
 &  &  &  &  Scd Spirals &  &  &  &  \\
NGC275 & 0.851 & -7.066 & -22.04 & 9.88 & 24.2 & 5.04 & 37 & 105 \\
NGC598 (M33) & 1.564 & 30.660 & -20.95 & 4.10 & 1.0 & 9.93 & 54 & 23 \\
NGC1494 & 3.962 & -48.908 & -21.15 & 9.82 & 15.6 & 7.90 & 64 & 1 \\
NGC2280 & 6.747 & -27.639 & -23.86 & 8.26 & 26.4 & 25.94 & 64 & 164 \\
NGC2283 & 6.765 & -18.211 & -21.46 & 8.83 & 11.4 & 6.33 & 43 & 175 \\
NGC2403 & 7.614 & 65.603 & -21.56 & 6.19 & 3.6 & 12.09 & 60 & 126 \\
NGC2541 & 8.244 & 49.061 & -19.36 & 10.09 & 7.8 & 6.55 & 62 & 167 \\
NGC3184 & 10.305 & 41.424 & -22.35 & 7.23 & 8.2 & 9.38 & 17 &  \\
NGC3319 & 10.653 & 41.687 & -19.99 & 10.07 & 10.3 & 8.75 & 59 & 37 \\
NGC4654 & 12.732 & 13.126 & -23.05 & 7.74 & 14.4 & 10.49 & 58 & 122 \\
NGC5457 (M101) & 14.053 & 54.348 & -25.40 & 5.51 & 7.2 & 31.98 & 8 & 110\tablenotemark{d} \\
NGC5474 & 14.084 & 53.662 & -16.89 & 9.48 & 1.9 & 1.21 & 42 & 79 \\
NGC6140 & 16.349 & 65.390 & -20.76 & 9.74 & 12.6 & 8.95 & 37 & 76 \\
NGC6946 & 20.581 & 60.154 & -23.40 & 5.37 & 5.7 & 9.37 & 30 & 62\tablenotemark{d} \\
NGC7418 & 22.943 & -37.030 & -22.99 & 8.52 & 20.1 & 11.45 & 41 & 139 \\
NGC7424 & 22.955 & -41.071 & -21.33 & 9.25 & 13.1 & 14.88 & 41 & 115\tablenotemark{e}\\
 &  &  &  &  Sd Spirals &  &  &  &  \\
NGC300 & 0.915 & -37.685 & -20.66 & 6.38 & 2.6 & 7.50 & 48 & 114 \\
NGC337 & 0.997 & -7.578 & -22.70 & 9.10 & 22.9 & 9.61 & 53 & 60 \\
NGC925 & 2.455 & 33.579 & -21.56 & 7.87 & 7.7 & 12.60 & 57 & 102 \\
NGC1313 & 3.304 & -66.497 & -21.46 & 7.57 & 6.4 & 8.60 & 40 & 39 \\
NGC4145 & 12.167 & 39.884 & -22.26 & 8.48 & 14.1 & 11.67 & 55 & 100 \\
NGC4519 & 12.558 & 8.655 & -21.59 & 9.56 & 17.0 & 7.07 & 37 & 149 \\
NGC5068 & 13.315 & -21.039 & -22.30 & 7.55 & 9.3 & 10.11 & 26 & 110\tablenotemark{d} \\
NGC5585 & 14.330 & 56.729 & -19.89 & 9.50 & 7.5 & 6.06 & 53 & 32 \\
NGC5556 & 14.343 & -29.242 & -21.86 & 9.55 & 19.2 & 11.82 & 48 & 143 \\
NGC7320 & 22.601 & 33.948 & -19.64 & 10.52 & 15.3 & 4.25 & 55 & 133 \\
NGC7793 & 23.964 & -32.591 & -21.15 & 6.86 & 4.0 & 5.91 & 58 & 84 \\
\tableline
\enddata
\tablenotetext{a}{HyperLeda Extragalactic Database (LEDA)}
\tablenotetext{b}{2MASS search engine GATOR}
\tablenotetext{c}{NED}
\tablenotetext{d}{\citealt{martin95}}
\tablenotetext{e}{2MASS IRSA}
\tablenotetext{f}{\citealt{heraudeau96}}

\end{deluxetable*}

\begin{deluxetable*}{rrcccccc}
\tabletypesize{\scriptsize}
\tablecaption{2MASS Bars\label{resultstab}}
\tablehead{
\colhead{Galaxy} & \colhead{Type\tablenotemark{a}} & \colhead{$\epsilon_{max}$} & \colhead{PA$_{bar}$} & \colhead{a $\pm \delta a$\tablenotemark{b}} & \colhead{a$_{deproj}$} & \colhead{$a_{bar}$} & \colhead{$a_{bar}/R_{25}$} \\
\colhead{} & \colhead{(RC3)} & \colhead{} & \colhead{($\deg$)} & \colhead{($\arcsec$)} & \colhead{($\arcsec$)} & \colhead{(kpc)} & \colhead{}}
\tablenotetext{a}{as given in NED}
\tablenotetext{b}{error in measurement, see \S\ref{signature}}
\startdata
&  &  &  Barred &  &  &  &  \\
\tableline
& & & S0/a Spirals & & & & \\
NGC1291 & SB0/a & 0.40 & 171 & 87 $\pm$ 5 & 88 & 5.0 & 0.29 \\
NGC1317 & SAB(rl)0/a & 0.24 & 150 & 42 $\pm$ 2 & 49 & 6.3 & 0.54 \\
NGC1326 & SB(rl)0/a & 0.37 & 20 & 30 $\pm$ 3 & 39 & 3.6 & 0.30 \\
NGC2217 & SB(rs)0/a & 0.43 & 113 & 39 $\pm$ 2 & 43 & 4.7 & 0.29 \\
NGC2681 & SAB(rs)0/a;Sy & 0.21 & 77 & 18 $\pm$ 2 & 18 & 0.8 & 0.16 \\
NGC2655 & SAB(s)0/a & 0.32 & 85 & 33 $\pm$ 5 & 53 & 5.0 & 0.37 \\
NGC5101 & SB(rl)0/a & 0.51 & 122 & 48 $\pm$ 3 & 48 & 6.0 & 0.28 \\
 &  &  &  Sa Spirals &  &  &  &  \\
NGC3718 & SB(s)a;pec;Sy1 & 0.20 & 12 & 60 $\pm$ 3 & 65 & 4.3 & 0.29 \\
NGC4314 & SB(rs)a & 0.63 & 148 & 69 $\pm$ 8 & 69 & 4.6 & 0.59 \\
NGC4491 & SB(s)a: & 0.54 & 138 & 24 $\pm$ 3 & 25 & 0.8 & 0.52 \\
NGC7172 & Sa\_pec\_Sy2 & 0.51 & 96 & 45 $\pm$ 5 & 46 & 8.0 & 0.68 \\
NGC7727 & SAB(s)a\_pec & 0.21 & 87 & 27 $\pm$ 2 & 33 & 4.2 & 0.27 \\
 &  &  &  Sab Spirals &  &  &  &  \\
NGC1350 & SB(r)ab & 0.55 & 36 & 51 $\pm$ 5 & 68 & 8.3 & 0.39 \\
NGC1398 & SB(rs)ab;Sy & 0.33 & 12 & 39 $\pm$ 2 & 55 & 5.2 & 0.25 \\
NGC1433 & SB(rs)ab\_Sy2 & 0.62 & 96 & 90 $\pm$ 8 & 90 & 6.5 & 0.48 \\
NGC1512 & SB(r)ab & 0.64 & 44 & 72 $\pm$ 3 & 77 & 4.7 & 0.25 \\
NGC4450 & SA(s)ab & 0.50 & 6 & 42 $\pm$ 3 & 43 & 5.7 & 0.28 \\
NGC4725 & SAB(r)ab;pec & 0.65 & 48 & 132 $\pm$ 9 & 139 & 11.2 & 0.43 \\
NGC4941 & SAB(r)ab\_Sy2 & 0.53 & 16 & 93 $\pm$ 6 & 94 & 7.1 & 0.88 \\
NGC5317 & SA(rs)bc;pec & 0.43 & 47 & 36 $\pm$ 3 & 37 & 2.9 & 0.76 \\
NGC7552 & SB(s)ab & 0.59 & 96 & 48 $\pm$ 3 & 48 & 5.2 & 0.43 \\
 &  &  &  Sb Spirals &  &  &  &  \\
NGC1097 & SBb;Sy1 & 0.61 & 147 & 87 $\pm$ 6 & 88 & 7.5 & 0.29 \\
NGC1365 & SBb(s)b;Sy1.8 & 0.59 & 85 & 90 $\pm$ 5 & 152 & 13.3 & 0.45 \\
NGC3227 & SAB(s);pec;Sy & 0.59 & 151 & 57 $\pm$ 8 & 58 & 4.5 & 0.39 \\
NGC772 & SA(s)b & 0.24 & 116 & 21 $\pm$ 3 & 23 & 3.8 & 0.11 \\
NGC986 & SB(rs)b & 0.63 & 58 & 48 $\pm$ 6 & 61 & 8.1 & 0.51 \\
NGC1068 & SA(rs)b;Sy1;2 & 0.33 & 48 & 15 $\pm$ 2 & 15 & 1.2 & 0.07 \\
NGC1530 & SB(rs)b & 0.61 & 124 & 51 $\pm$ 2 & 77 & 14.0 & 0.62 \\
NGC3351 & SB(r)b;HII & 0.43 & 113 & 57 $\pm$ 3 & 103 & 5.4 & 0.45 \\
NGC3583 & SB(s)b & 0.40 & 76 & 21 $\pm$ 2 & 33 & 4.8 & 0.47 \\
NGC3673 & SB(r)b & 0.68 & 83 & 45 $\pm$ 3 & 48 & 6.3 & 0.42 \\
NGC4548 (M91) & SBb(rs);Sy & 0.49 & 60 & 57 $\pm$ 2 & 70 & 2.3 & 0.45 \\
NGC4579 & SAB(rs)b;Sy1.9 & 0.46 & 59 & 39 $\pm$ 3 & 42 & 4.3 & 0.25 \\
NGC4593 & SB(rs)b\_Sy1 & 0.59 & 58 & 48 $\pm$ 5 & 53 & 9.2 & 0.60 \\
NGC5383 & SB(rs)b:pec & 0.57 & 127 & 48 $\pm$ 5 & 50 & 8.0 & 0.63 \\
NGC5850 & SB(r)b & 0.60 & 117 & 60 $\pm$ 3 & 61 & 10.5 & 0.47 \\
NGC5985 & SAB(r)b;Sy1 & 0.59 & 15 & 48 $\pm$ 3 & 48 & 8.1 & 0.31 \\
NGC6300 & SB(rs)b & 0.45 & 71 & 36 $\pm$ 5 & 52 & 3.9 & 0.34 \\
 &  &  &  Sbc Spirals &  &  &  &  \\
NGC289 & SAB(rs)bc & 0.48 & 122 & 21 $\pm$ 2 & 21 & 2.3 & 0.13 \\
NGC613 & SB(rs)bc & 0.70 & 124 & 75 $\pm$ 9 & 75 & 7.5 & 0.46 \\
NGC1300 & SB(s)bc & 0.75 & 103 & 84 $\pm$ 9 & 84 & 8.7 & 0.45 \\
NGC1566 & SAB(rs)bc;Sy1 & 0.38 & 3 & 33 $\pm$ 2 & 36 & 3.7 & 0.14 \\
NGC1672 & SB(r)bc\_Sy2 & 0.63 & 96 & 72 $\pm$ 6 & 102 & 7.7 & 0.50 \\
NGC2559 & SB(s)bc\_pec: & 0.53 & 39 & 27 $\pm$ 2 & 39 & 4.1 & 0.32 \\
NGC3344 & SAB(r)bc & 0.28 & 2 & 24 $\pm$ 2 & 25 & 1.0 & 0.11 \\
NGC3882 & SB(s)bc & 0.57 & 124 & 33 $\pm$ 2 & 36 & 4.4 & 0.47 \\
NGC3953 & SB(r)bc & 0.47 & 46 & 27 $\pm$ 2 & 40 & 2.9 & 0.18 \\
NGC3992 (M109) & SB(rs)bc;LINER & 0.60 & 39 & 51 $\pm$ 5 & 65 & 4.6 & 0.31 \\
NGC4051 & SAB(rs)bc & 0.65 & 134 & 63 $\pm$ 3 & 63 & 3.0 & 0.41 \\
NGC4303 (M61) & SAB(rs)bc;HII & 0.59 & 3 & 48 $\pm$ 3 & 48 & 5.1 & 0.26 \\
NGC4321 (M100) & SAB(s)bc;LINER & 0.51 & 106 & 57 $\pm$ 3 & 66 & 7.0 & 0.29 \\
NGC4639 & SAB(rs)bc & 0.47 & 169 & 24 $\pm$ 3 & 30 & 2.0 & 0.33 \\
NGC4930 & SB(rs)bc & 0.56 & 43 & 42 $\pm$ 2 & 42 & 7.3 & 0.30 \\
NGC5054 & SA(s)bc & 0.31 & 162 & 21 $\pm$ 2 & 21 & 2.5 & 0.15 \\
NGC5194 (M51A) & SA(s)bc & 0.27 & 139 & 15 $\pm$ 2 & 16 & 0.5 & 0.06 \\
NGC5371 & SAB(rs)bc & 0.30 & 94 & 24 $\pm$ 5 & 36 & 6.1 & 0.29 \\
NGC5347 & SB(rs)ab\_Sy2 & 0.53 & 103 & 30 $\pm$ 2 & 33 & 2.8 & 0.18 \\
NGC5713 & SAB(rs)bc;pec & 0.45 & 104 & 18 $\pm$ 2 & 23 & 3.0 & 0.29 \\
NGC5921 & SB(r)bc & 0.58 & 16 & 39 $\pm$ 3 & 50 & 5.0 & 0.38 \\
NGC6221 & SB(s)bc;pec;Sy2 & 0.43 & 117 & 24 $\pm$ 2 & 47 & 4.7 & 0.35 \\
NGC6384 & SAB(r)bc & 0.50 & 35 & 24 $\pm$ 2 & 28 & 3.2 & 0.19 \\
NGC6744 & SAB(r)bc & 0.68 & 179 & 90 $\pm$ 8 & 95 & 5.4 & 0.15 \\
\enddata

\end{deluxetable*}

\setcounter{table}{1}
\begin{deluxetable*}{rrcccccc}
\tabletypesize{\scriptsize}
\tablecaption{2MASS Bars $-$ Continued}
\tablehead{
\colhead{Galaxy} & \colhead{Type\tablenotemark{a}} & \colhead{$\epsilon_{max}$} & \colhead{PA$_{bar}$} & \colhead{a $\pm \delta a$\tablenotemark{b}} & \colhead{a$_{deproj}$} & \colhead{$a_{bar}$} & \colhead{$a_{bar}/R_{25}$} \\
\colhead{} & \colhead{(RC3)} & \colhead{} & \colhead{($\deg$)} & \colhead{($\arcsec$)} & \colhead{($\arcsec$)} & \colhead{(kpc)} & \colhead{}}
\tablenotetext{a}{as given in NED}
\tablenotetext{b}{error in measurement, see \S\ref{signature}}
\startdata
&  &  &  Barred &  &  &  &  \\
\tableline
 &  &  &  Sc Spirals &  &  &  &  \\
NGC578 & SAB(rs)c & 0.53 & 84 & 18 $\pm$ 2 & 22 & 2.4 & 0.16 \\
NGC864 & SAB(rs)c & 0.44 & 102 & 24 $\pm$ 2 & 35 & 3.7 & 0.26 \\
NGC1073 & SB(rs)c & 0.66 & 59 & 45 $\pm$ 3 & 48 & 3.9 & 0.34 \\
NGC1187 & SB(r)c & 0.56 & 134 & 36 $\pm$ 3 & 36 & 3.4 & 0.22 \\
NGC1232 & SAB(rs)c & 0.42 & 86 & 21 $\pm$ 3 & 21 & 2.4 & 0.10 \\
NGC1637 & SAB(rs)c & 0.43 & 68 & 24 $\pm$ 3 & 27 & 1.3 & 0.28 \\
NGC2835 & SAB(rs)c & 0.41 & 115 & 21 $\pm$ 2 & 31 & 1.9 & 0.15 \\
NGC3359 & SB(rs)c & 0.68 & 7 & 42 $\pm$ 5 & 43 & 2.9 & 0.20 \\
NGC3486 & SAB(r)c & 0.39 & 73 & 21 $\pm$ 2 & 21 & 1.0 & 0.10 \\
NGC3614 & SAB(r)c & 0.39 & 87 & 27 $\pm$ 2 & 27 & 4.2 & 0.22 \\
NGC3631 & SA(s)c & 0.40 & 13 & 51 $\pm$ 6 & 61 & 4.8 & 0.42 \\
NGC3726 & SAB(r)c & 0.70 & 31 & 39 $\pm$ 3 & 41 & 2.4 & 0.23 \\
NGC4535 & SAB(s)c & 0.62 & 37 & 45 $\pm$ 9 & 51 & 6.7 & 0.25 \\
NGC4647 & SAB(rs)c & 0.21 & 88 & 21 $\pm$ 2 & 23 & 2.2 & 0.27 \\
NGC5236 (M83) & SAB(s)c;Sbrst & 0.56 & 58 & 114 $\pm$ 5 & 114 & 4.0 & 0.27 \\
NGC5643 & SAB(rs)c & 0.65 & 84 & 54 $\pm$ 5 & 54 & 4.4 & 0.36 \\
 &  &  &  Scd Spirals &  &  &  &  \\
NGC2283 & SB(s)cd & 0.55 & 173 & 12 $\pm$ 2 & 12 & 0.7 & 0.11 \\
NGC2403 & SAB(s)cd & 0.41 & 113 & 15 $\pm$ 2 & 16 & 0.3 & 0.02 \\
NGC5457 (M101) & SAB(rs)cd & 0.45 & 82 & 51 $\pm$ 2 & 51 & 1.8 & 0.06 \\
NGC6946 & SAB(rs)cd & 0.46 & 17 & 60 $\pm$ 2 & 65 & 1.8 & 0.19 \\
NGC7418 & SAB(rs)cd & 0.61 & 126 & 30 $\pm$ 2 & 31 & 3.0 & 0.26 \\
NGC7424 & SAB(rs)cd & 0.51 & 130 & 18 $\pm$ 3 & 18 & 1.2 & 0.08 \\
 &  &  &  Sd Spirals &  &  &  &  \\
NGC337 & SB(s)d & 0.45 & 162 & 21 $\pm$ 6 & 35 & 3.9 & 0.40 \\
NGC1313 & SB(s)d & 0.59 & 15 & 33 $\pm$ 9 & 35 & 1.1 & 0.13 \\
NGC4145 & SAB(rs)d;LINER & 0.34 & 135 & 12 $\pm$ 3 & 16 & 1.1 & 0.09 \\
NGC5068 & SB(s)d & 0.55 & 149 & 21 $\pm$ 3 & 22 & 1.0 & 0.10 \\
NGC5556 & SAB(rs)d & 0.48 & 93 & 15 $\pm$ 2 & 20 & 1.8 & 0.16 \\
\tableline
& & & Barred Candidates & & & & \\
\tableline
& & &  Sa Spirals & & & & \\
NGC1022 & SB(s)a;Sbrst & 0.42 & 108 & 18 $\pm$ 2 & 22 & 2.1 & 0.28 \\
NGC1367 & SABa & 0.39 & 104 & 24 $\pm$ 2 & 28 & 2.8 & 0.16 \\
NGC5728 & SAB(r)a;Sy2 & 0.66 & 33 & 54 $\pm$ 3 & 72 & 13.0 & 0.76 \\
 &  &  &  Sab Spirals &  &  &  &  \\
NGC4151 & SAB(rs)ab & 0.48 & 132 & 63 $\pm$ 5 & 82 & 5.5 & 0.45 \\
 &  &  &  Sb Spirals &  &  &  &  \\
NGC210 & SAB(s)b & 0.47 & 173 & 33 $\pm$ 3 & 34 & 3.7 & 0.24 \\
NGC7606 & SA(s)b & 0.61 & 142 & 36 $\pm$ 5 & 36 & 5.4 & 0.28 \\
 &  &  &  Sbc Spirals &  &  &  &  \\
NGC2207 & SAB(rs)bc\_pec & 0.42 & 64 & 24 $\pm$ 3 & 42 & 7.7 & 0.30 \\
NGC4567 (VV219B) & SA(rs)bc & 0.45 & 58 & 30 $\pm$ 6 & 33 & 5.1 & 0.40 \\
 &  &  &  Sc Spirals &  &  &  &  \\
NGC2976 & SAc;pec & 0.69 & 137 & 72 $\pm$ 5 & 73 & 1.4 & 0.42 \\
 &  &  &  Scd Spirals &  &  &  &  \\
NGC275 & SB(rs)cd\_pec & 0.35 & 78 & 12 $\pm$ 3 & 13 & 1.5 & 0.30 \\
NGC1494 & SAB(rs)cd & 0.47 & 174 & 15 $\pm$ 3 & 16 & 1.2 & 0.15 \\
NGC2280 & SA(s)cd & 0.60 & 153 & 48 $\pm$ 3 & 52 & 6.7 & 0.26 \\
\tableline
\enddata

\end{deluxetable*}

\begin{deluxetable*}{lccc}
\tabletypesize{\scriptsize}
\tablecaption{Local Bar Properties \label{TypBartab}}
\tablehead{
\colhead{Property\tablenotemark{a}} & \colhead{2MASS Bar}  & \colhead{Early-Type 2MASS Bar}  & \colhead{Late-Type 2MASS Bar}  \\
\colhead{} & \colhead{(S0/a-Sd)}  & \colhead{(Sa-Sb)}  & \colhead{(Sc-Sd)} 
}
\startdata
\tableline
$a_{bar}$ (kpc) & 4.2 $\pm$2.9 & 5.4 $\pm$3.3 & 2.2 $\pm$1.7 \\
$a_{bar}$ / $R_{25}$ & 0.29 $\pm$0.17 & 0.43 $\pm$0.18 & 0.22 $\pm$0.11 \\
$\epsilon_{bar}$ & 0.50 $\pm$0.13 & 0.54 $\pm$0.13 & 0.48 $\pm$0.12 \\
\tableline
\tablenotetext{a}{Median value $\pm 1 \sigma$ standard deviation}
\enddata

\end{deluxetable*}


\begin{thebibliography}

\bibitem[Abraham et al.(1996)]{abraham96}Abraham, R. G., Tanvir, N. R., Santiago, B. X., Ellis, R. S., Glazebrook, K., van den Bergh, S. 1996, \mnras, 279, 47

\bibitem[Abraham et al.(1999)]{abraham99}Abraham, R. G., Merrifield, M. R., Ellis, R. S., Tanvir, N. R., Brinchmann, J. 1999, \mnras, 308, 569

\bibitem[Abraham et al.(2000)]{abraham00}Abraham, R. G., Merrifield, M. R. 2000, \aj, 120, 2835

\bibitem[Athanassoula(1983)]{a83} Athanassoula, E., Bienaym\'{e}, O., Martinet, L., Pfenniger, D. 1983, \aap, 127, 349

\bibitem[Athanassoula(1992a)]{a92a} Athanassoula, E. 1992a, \mnras, 259, 328

\bibitem[Athanassoula(1992b)]{a92b} Athanassoula, E. 1992b, \mnras, 259, 345

\bibitem[Athanassoula(2005)]{a05} Athanassoula, E., Lambert, J. C., Dehnen, W. 2005, \mnras, 363, 496

\bibitem[Block et al.(2004)]{block04} Block, D., Buta, R., Knapen, J.H., Elmegreen, B.G., Elmegreen, D.M., \& Puerari, I. 2004, \aj, 128, 183

\bibitem[Bournaud \& Combes(2002)]{bournaud02} Bournaud, F., \&
Combes, F. 2002, \aap, 392, 83

\bibitem[Buta \& Block(2001)]{buta01} Buta, R., \& Block, D. L., 2001,
\apj, 550, 243

\bibitem[Buta et al.(2005)]{buta05} Buta, R., Vasylyev, S., Salo, H., Laurikainen, E., 2005,
\aj, 130, 506

\bibitem[Combes \& Gerin(1985)]{combes85} Combes, F., \& Gerin, M. 1985, \aap, 150, 327

\bibitem[Das et al.(2003)]{das03} Das, M., Teuben, P. J., Vogel, S. N., Regan, M. W., Sheth, K. , Harris, A. I., Jeffreys, W. H., 2003, \apj, 582, 190

\bibitem[de Vaucouleurs et al.(1991)]{devau91} de Vaucouleurs, G., de
Vaucouleurs, A., Corwin, H. G., Jr., Buta, R. J., Paturel, G., \&
Fouque, P. 1991, Third Reference Catalogue of Bright Galaxies, (New
York:Springer-Verlag)

\bibitem[de Vaucouleurs(1963)]{devau63} de Vaucouleurs, G. 1963, \apjs, 8, 31D 

\bibitem[Dickinson \& Giavalisco(2002)]{dickinson01} Dickinson, M., Giaalvisco, M. \& the GOODS team in the proceedings of the ESO/USM workshp "The Mass of Galaxies at Low and High Redshift", (Venice, Italy), 2001 eds. Bender, R. \& Renzini, A.

\bibitem[Elmegreen \& Elmegreen(1985)]{elmegreen85} Elmegreen, B. G., \&
Elmegreen, D.M. 1985, \apj, 288, 438

\bibitem[Elmegreen et al.(2004)]{elmegreen04} Elmegreen, B. G.,
Elmegreen, D. M., Hirst, A. C. 2004, \apj, 612, 191

\bibitem[Elmegreen et al.(2005)]{elmegreen05} Elmegreen, D. M., Elmegreen, B. G., Rubin, D. S., Schaffer, M. A. 2005, \apj, 631, 85

\bibitem[Eskridge et al.(2000)]{eskridge00} Eskridge, P., et al. 2000, \aj, 119, 536 (E2000)

\bibitem[Eskridge et al.(2002)]{eskridge02} Eskridge, P. et al. 2002, \apjs, 143, 73 

\bibitem[Erwin(2005)]{erwin05} Erwin, P. 2005, \mnras, 364, 283

\bibitem[Friedli \& Benz(1993)]{friedli93} Friedli, D. \& Benz,
W. 1993, \aap, 268, 65

\bibitem[Friedli \& Pfenninger(1991)]{friedli91} Friedli, D., Pfenninger, D. 1991, IAUS, 146, 362

\bibitem[Hackwell \& Schweizer(1983)]{hackwell83}Hackwell, J. A.,
Schweizer, F., 1983, \apj, 265, 643

\bibitem[Jedrzejewski(1987)]{jedrzejewski87} Jedrzejewski, R. I. 1987,
\mnras, 226, 747

\bibitem[Jogee et al.(2004)]{jogee04} Jogee, S. et al. 2004, \apj, 615, 105

\bibitem[Heraudeau  \& Simien(1996)]{heraudeau96} Heraudeau, P., \& Simien, F.. 1996, \aap, 118, 111

\bibitem[Ho, Filippenko, \& Sargent(1997a)]{ho97a}Ho, L. C., Filippenko,
A. V., \& Sargent, W. L. W. 1997a, \apjs, 112, 315

\bibitem[Ho, Filippenko, \& Sargent(1997b)]{ho97b}Ho, L. C., Filippenko,
A. V., \& Sargent, W. L. W. 1997b, \apj, 487, 591

\bibitem[Jarrett et al.(2003)]{jarrett03}Jarrett, T. H., Chester, T., Cutri, R., Schneider, S. E., Huchra, J. P. 2003, AJ, 125, 525

\bibitem[Knapen et al.(2000)]{knapen00} Knapen, J., Shlosman, I., Peletier, R. 2000, \apj, 529, 93

\bibitem[Kormendy(1979)]{kormendy79} Kormendy, J. 1979, \apj, 227, 714

\bibitem[Kormendy(1982)]{kormendy82} Kormendy, J. 1982, \apj, 257, 75

\bibitem[Kormendy \& Kennicutt(2004)]{kormendy04} Kormendy, J. \& Kennicutt, R.C. 2004, \araa, 42, 603

\bibitem[Laine et al.(2002)]{laine02} Laine, S., Shlosman, I., Knapen,
J. H., \& Peletier, R. F. 2002, \apj, 567, 97

\bibitem[Laurikainen, Salo \& Rautiainen (2002)]{laurikainen02} Laurikainen,
E., Salo, H., \& Rautiainen, P. 2002, \mnras, 331, 880

\bibitem[Laurikainen, Salo \& Buta (2004)]{laurikainen04} Laurikainen,
E., Salo, H., \& Buta, R. 2004, \apj, 607, 103 (LSB04)

\bibitem[Martin (1995)]{martin95} Martin, P. 1995, \aj, 109, 2428

\bibitem[Martin \& Roy(1994)]{martin94} Martin, P., \& Roy, J. 1994,
\apj, 424, 599

\bibitem[Mulchaey \& Regan(1997)]{mulchaey97} Mulchaey, J. S., \&
Regan, M. W. 1997, \apjl, 482, 135

\bibitem[Norman, Sellwood, \& Hasan(1996)]{norman96} Norman, C. A.,
Sellwood, J. A., \& Hasan, H. 1996, \apj, 462, 114

\bibitem[Ostriker \& Peebles (1973)]{ostriker73} Ostriker, J. P., Peebles,
P. J. E. 1973, \apj, 186, 467

\bibitem[Piner, Stone \& Teuben(1995)]{piner95} Piner, B.G., Stone,
J.M., \& Teuben, P. J. 1995, \apj, 449, 508

\bibitem[Regan \& Elmegreen(1997)]{regan97a} Regan, M. W., Elmegreen, D. M. 1997, \apj, 114, 965

\bibitem[Regan, Vogel \& Teuben(1997)]{regan97b} Regan, M. W., Vogel,
S.N., Teuben, P. J. 1997, \apjl, 482, 135

\bibitem[Regan, Sheth \& Vogel(1999)]{regan99} Regan, M. W., Sheth,
K., \& Vogel, S.N. 1999, \apj, 526, 97

\bibitem[Rix et al.(2004)]{rix04} Rix, H-W., et al. 2004, \apjs, 152, 163

\bibitem[Sakamoto et al.(1999)]{sakamoto99} Sakamoto, K., Okumura,
S. K., Ishizuki, S., Scoville, N. Z. 1999b, \apj, 525, 691

\bibitem[Scoville et al. (1988)]{scoville88} Scoville, N. Z., Matthews, K.,
Carico, D. P., Sanders, D. B. 1988, \apj, 327, 61

\bibitem[Scoville et al.(2006)]{scoville06} Scoville, N. Z., \& COSMOS team, \apj, submitted

\bibitem[Seigar \& James(1998)]{seigar98}Seigar, M. S., James,
P. A. 1998, \mnras, 299, 685

\bibitem[Shen \& Sellwood(2004)]{shen04} Shen, J., Sellwood, J. A. 2004, \apj, 604, 614

\bibitem[Sheth et al.(2000)]{sheth00} Sheth, K., Regan, M. W., Vogel,
S. N., \& Teuben, P. J. 2000 \apj, 532, 221

\bibitem[Sheth et al.(2002)]{sheth02} Sheth, K., Vogel, S. N., Regan,
M. W., Teuben, P.J., Harris, A. I., \& Thornley, M. D. 2002, \aj, 124, 2581

\bibitem[Sheth et al.(2003)]{sheth03} Sheth, K., Regan,
M. W., Scoville, N.Z., \& Strubbe, L.E. 2003, \apjl, 592, 13

\bibitem[Sheth et al.(2004)]{sheth04} Sheth, K., Men\'{e}ndez-Delmestre, K., Scoville, N., Jarrett, T., Strubbe, L., Regan, M. W., Schinnerer, E., Block, D. L. 2004, Penetrating Bars Through Masks of Cosmic Dust, eds. Block, D. L., Puerari, I., Freeman, K. C., Groess, R., Block, E. K., Astrophysics and Space Science Library 319, p405

\bibitem[Sheth et al.(2005)]{sheth05} Sheth, K., Vogel, S.N., Regan, M.W., Thornley, M.D., \& Teuben, P.J. 2005, \apj, 632, 217

\bibitem[Shlosman, Frank, \& Begelman(1989)]{shlosman89} Shlosman, I.,
Frank, J., \& Begelman, M. C. 1989, \nat, 338, 45

\bibitem[Skrutskie et al.(2006)]{skrutskie06} Skrutskie, M. F. et al. 2006, \aj, 131, 1163

\bibitem[Teuben(1995)]{teuben95} Teuben, P. J. 1995, The Stellar
Dynamics Toolbox NEMO, in: Astronomical Data Analysis Software and
Systems IV, eds. Shaw, R., Payne, H.E. \& Hayes, J.J.E., PASP Conf
Series 77, p398.

\bibitem[Thronson et al.(1989)]{thronson89} Thronson, H. A. et
al. 1989, \apj, 343, 158

\bibitem[van den Bergh et al.(1996)]{vandenBergh96}van den Bergh, S., Abraham, R. G., Ellis, R., Tanvir, N. R., Santiago, B. X., Glazebrook, K. G. 1996, \apj, 112, 359

\bibitem[van den Bergh et al.(2000)]{vandenBergh00}van den Bergh, S., Cohen, J. G., Hogg, D. W., Blandford, R. 2000, \apj, 120, 2190

\bibitem[van den Bergh et al.(2001)]{vandenBergh01}van den Bergh, S., Cohen, J. G., Crabbe, C. 2001, \apj, 122, 611

\bibitem[Whyte et al.(2002)]{whyte02} Whyte, L. F., Abraham, R. G., Merrifield, M. R., Eskridge, P. B., Frogel, J. A., Pogge, R. W. 2002, \mnras, 336, 1281 (W2002)

\bibitem[Wozniak et al.(1991)]{wozniak91} Wozniak, H., Pierce, M. J. 1991, \aap, 88, 325

\bibitem[Zheng et al.(2005)]{zheng05} Zheng, X. Z., Hammer, F., Flores, H., Ass\'{e}mat, F., Rawat, A. 2005, \aap, 435, 507

\end{thebibliography}
\end{document}